\documentclass[10pt,journal,compsoc,cspaper,twoside]{IEEEtran}

\newcommand{\subparagraph}{}

\usepackage{booktabs} 

\ifCLASSOPTIONcompsoc
  \usepackage[nocompress]{cite}
\else
  \usepackage{cite}
\fi

\ifCLASSINFOpdf
  \usepackage[pdftex]{graphicx}
  \DeclareGraphicsExtensions{.pdf,.jpeg,.png}
\else
  \usepackage[dvips]{graphicx}
  \DeclareGraphicsExtensions{.eps}
\fi

\usepackage{amssymb}
\usepackage{amsmath}
\usepackage{algorithm}
\usepackage{algorithmic}

\usepackage{array}
\usepackage[caption=false,font=footnotesize]{subfig}
\usepackage{dblfloatfix}
\usepackage{url}
\usepackage{enumitem}
\usepackage{ulem}
\usepackage{nonfloat}
\DeclareMathOperator*{\argmax}{arg\,max}

\usepackage{color}

\usepackage[compact]{titlesec}
\usepackage[format=plain,labelfont=bf]{caption}

\newcommand{\zyj}[1]{\textcolor{black}{#1}}

\begin{document}
\title{Color Contrast Enhanced Rendering for Optical See-through Head-mounted Displays}



\author{Yunjin~Zhang,
        Rui~Wang,
        Yifan~(Evan)~Peng,
        Wei~Hua,
        and~Hujun~Bao%
\IEEEcompsocitemizethanks{\IEEEcompsocthanksitem Yunjin~Zhang, Rui~Wang, Wei~Hua, and Hujun~Bao were in the State Key Laboratory of CAD\&CG, Zhejiang University, China. \newline
Yifan~(Evan)~Peng was in Electrical Engineering, Stanford University, USA.
\newline
Corresponding author: Rui Wang, rwang@cad.zju.edu.cn
\newline
\textbf{This work has been submitted to the IEEE for possible publication. Copyright may be transferred without notice, after which this version may no longer be accessible.}}}

\markboth{SUBMITTED TO IEEE TRANSACTIONS ON VISUALIZATION AND COMPUTER GRAPHICS}%
{Zhang \MakeLowercase{\textit{et al.}}: Color Contrast Enhanced Rendering for Optical See-through Head-mounted Displays}



\IEEEtitleabstractindextext{

  \begin{center}
    \captionsetup{type=figure}
    \includegraphics[width=0.925\textwidth]{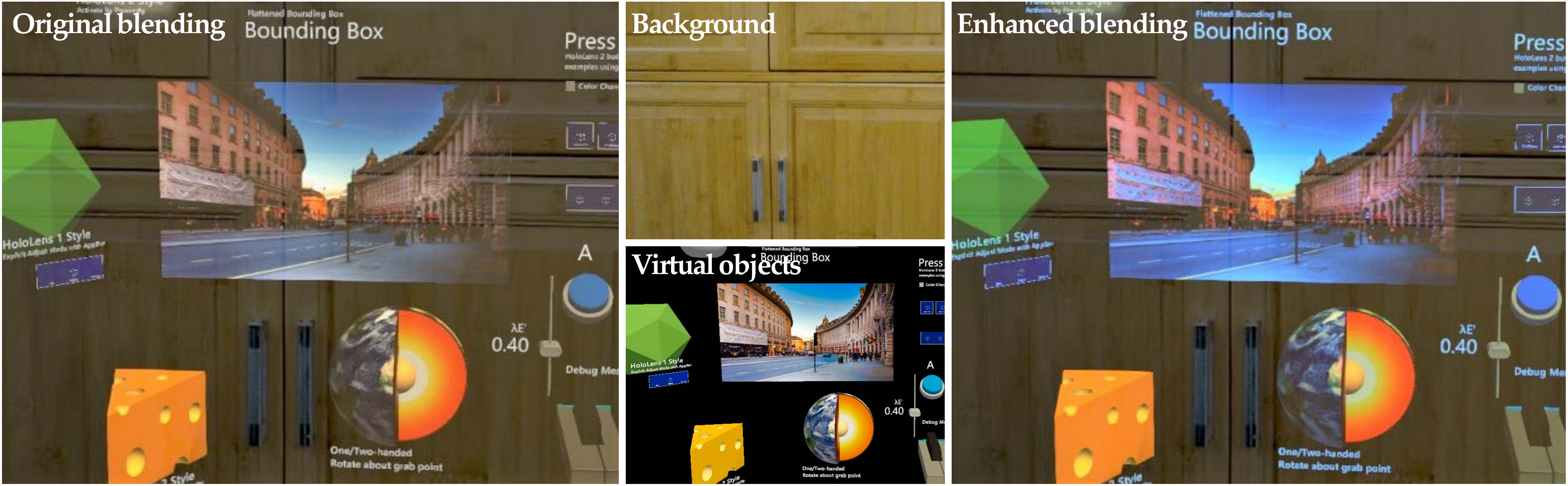}
    \captionof{figure}{Representative results of color contrast enhanced rendering for optical see-through head-mounted displays (OST-HMDs). \textbf{Left:} Original blending scene perceived from a commercially available OST-HMD in front of a typical background. \textbf{Middle:} The original background and virtual objects. \textbf{Right:} Our method improves the visual distinctions between rendered images and the physical environment by performing a constrained color optimization regarding the perception in chromaticity and luminance of the displayed color.}
    \setcounter{figure}{0}
    \label{fig:teaser}
 \end{center}

\begin{abstract}
Most commercially available optical see-through head-mounted displays (OST-HMDs) utilize optical combiners to simultaneously deliver the physical background and virtual objects to users. Given that the displayed images perceived by users are a blend of rendered pixels and background colors, high fidelity color perception in mixed reality (MR) scenarios using OST-HMDs is a challenge.
We propose a real-time rendering scheme to specifically enhance the color contrast between virtual objects and the surrounding background for OST-HMDs. Inspired by the discovery of color perception in psychophysics, we first formulate the color contrast enhancement as a constrained optimization problem. We then design an end-to-end algorithm to search the optimal complementary shift in both chromaticity and luminance of the displayed color. This aims at enhancing the contrast between virtual objects and real background as well as keeping the consistency with the original color. We assess the performance of our approach in a simulated environment and with a commercially available OST-HMD. Experimental results from objective evaluations and subjective user studies demonstrate that the proposed method makes rendered virtual objects more distinguishable from the surrounding background, thus bringing better visual experience.  
\end{abstract}


}

\maketitle




\IEEEraisesectionheading{\section{Introduction}\label{sec:introduction}}

\IEEEPARstart{I}{n} recent years, innovations in optical see-through head-mounted displays (OST-HMDs) have led to the rapid development of mixed reality (MR) technologies. In contrast to virtual reality (VR) HMDs or video see-through HMDs, OST-HMDs mainly enable users to perceive the real, non-rendered environment through the optics, as well as the virtual content through displays. This design scheme dramatically relieves the discomfort when wearing common video-based see-through HMDs. However, the optical combiner blends the rendered pixels with the physical background so that the virtual objects are unable to be presented independently. This essential property of combiners causes a color-blending problem, which hinders the ability of clearly observing the virtual content in MR scenarios, especially when the virtual and real scenes show low color contrast.

Existing solutions to the color-blending problem can be divided into two categories, namely, hardware- and software-based solutions. Hardware-based solutions physically adjust each pixel's transparency on displays to avoid the color blending of rendered images and the background~\cite{Cakmakci2004,Kiyokawa2000}. Although certain solutions exert every effort for miniaturization \cite{Gao2012,Wilson2017}, their scalability and flexibility are very limited on account of additional hardware. Software-based methods focus on color correction, seeking to modify the colors of rendered pixels to minimize the color blending of the virtual content and the background \cite{Weiland2009}. These approaches attempt to mitigate the color-blending effect via the subtraction compensation. Researchers have developed high-precision, pixel-precise color correction algorithms \cite{Hincapie-Ramos2015,Langlotz2016} and accurate colorimetric background estimation \cite{Kim2017}. However, for commercial OST-HMDs, such as Google Glass$\footnote{https://www.google.com/glass/}$, Microsoft HoloLens$\footnote{https://www.microsoft.com/en-us/hololens/}$, and Magic Leap$\footnote{https://www.magicleap.com/}$, using the subtraction compensation may decrease the brightness of the virtual content, resulting in low visual distinctions between rendered images and the physical environment. 

In certain MR scenarios, we note that the color correctness may not be the only purpose of displaying virtual objects. Instead, the color contrast of virtual objects to the real background should be sometimes pursued to allow users to better distinguish the perceived virtual objects from the background. To this point, intuitively increasing the brightness of the virtual objects may lead to a decreased contrast within their surfaces \cite{Fukiage2014}.
Therefore, this work proposes a novel real-time color contrast enhancement for OST-HMDs, aiming to improve the distinction between the rendered image and the real background, as well as to consider the consistency of enhanced colors to the original displayed colors. In particular, our work builds on the characteristic of the human visual system (HVS) that the color of one region induces the complementary color of the neighboring region in perception \cite{wolfe2006sensation}. 
We perform a constrained color optimization in the CIEL$^\ast$a$^\ast$b$^\ast$ color space to search the optimal complementary color under constraints regarding the perception in chromaticity and luminance.
Results show that the proposed method can enhance the perceived distinction between the virtual content and the surrounding background in typical environments. 

Particularly, our contributions are as follows:
\begin{itemize}
  \item We exploit a novel insight of color-blending problem that enhances color contrast to improve users' visual experience;
  \item We develop an end-to-end, real-time rendering algorithm to find optimal colors, improving the distinction between displayed images and physical environments on OST-HMDs;
  \item We demonstrate the effectiveness of our approach in a simulated environment and on a commercial OST-HMD.

\end{itemize}

\section{Related Work}
\label{sc:relatedwork}

MR devices and corresponding algorithms are becoming prolific research areas. Numerous studies explored how to better present colors of virtual scenes, such as harmonization \cite{Gruber2010}, defocus correction \cite{Itoh2015,Oshima2016}, color reproduction \cite{Menk2013,Itoh2015a}, color balancing \cite{Oskam2012}, light filtering \cite{Wetzstein2010,Mori2018,Itoh2019}, color correction \cite{Langlotz2016,Hincapie-Ramos2015}, and visibility improvement \cite{Fukiage2014, Lee2018}, etc. In this section, we introduce the most relevant research topics, color correction and visibility improvement, in more detail, and then describe a highly relevant work that the perceived color contrast in the HVS.

\textbf{Color Correction.} Solutions to color blending are commonly categorized into hardware- or software-based approaches. Hardware-based approaches commonly refer to occlusion support, which can avoid color blending physically by cutting rays off between background light and the user's eyes at the pixel level \cite{Kiyokawa2000}. Spatial-light modulators (SLM) provide the possibility to achieve this goal \cite{Cakmakci2004,Gao2012,Rathinavel2019}. By creating a mask pattern of transparency, OST-HMDs generate a black background in the mask area, blocking environment light and providing the occlusion effect. Certain methods use a high-speed switcher to frequently alternate between the virtual content and the real scene, achieving full or partial occlusion \cite{Maimone2013, Smithwick2014, Rhodes2019}. Their approaches, however, usually make the entire solutions impractical for daily use. An effort at compactness was made recently \cite{Wilson2017}, but its flexibility is often restricted owing to additional optical components.

By contrast, software-based methods attempt to mitigate color blending by changing the color of display pixels. Typical methods are known as color correction or compensation, which start from capturing the background and then accurately mapping the background image to the corresponding virtual content. Thereafter, the background color is subtracted from the rendered color of the virtual scene \cite{Weiland2009}. Certain research pays more attention to colorimetric estimation \cite{Ryu2016,Kim2017} or radiometric measurement \cite{Langlotz2016} of the background to obtain background information with higher accuracy. 

However, the aforementioned methods of subtraction compensation result in a decrease in the brightness of displayed images, thereby reducing the visibility of virtual content in the daily environment. One work manages to increase visibility with high contrast \cite{Hincapie-Ramos2015} but is demonstrated mainly for the textual content. In general, our goal in the current work is to improve the distinctions between virtual objects and physical backgrounds rather than just keep the perceived image consistent with the input, which is the primary difference between our method and color correction.

\textbf{Visibility Improvement.} Complementary to color correction, few recent works focus on improving the visibility of the virtual content. Fukiage et al. \cite{Fukiage2014} proposed an adaptive blending method on the basis of a subjective metric of visibility. However, on common OST-HMDs, this method increases the brightness but reduces the contrast of virtual objects, leading to a washed-out effect of the texture details. Lee and Kim \cite{Lee2018} used tone mapping to enhance the visibility of low gray-level regions under ambient light, but they only demonstrated on grayscale images. The main problem of these two approaches is that they only consider the luminance of rendered images, limiting their effectiveness for virtual objects with colorful, intricate textures.

\setcounter{figure}{1}
\begin{figure}
\centering 
\includegraphics[width=\columnwidth]{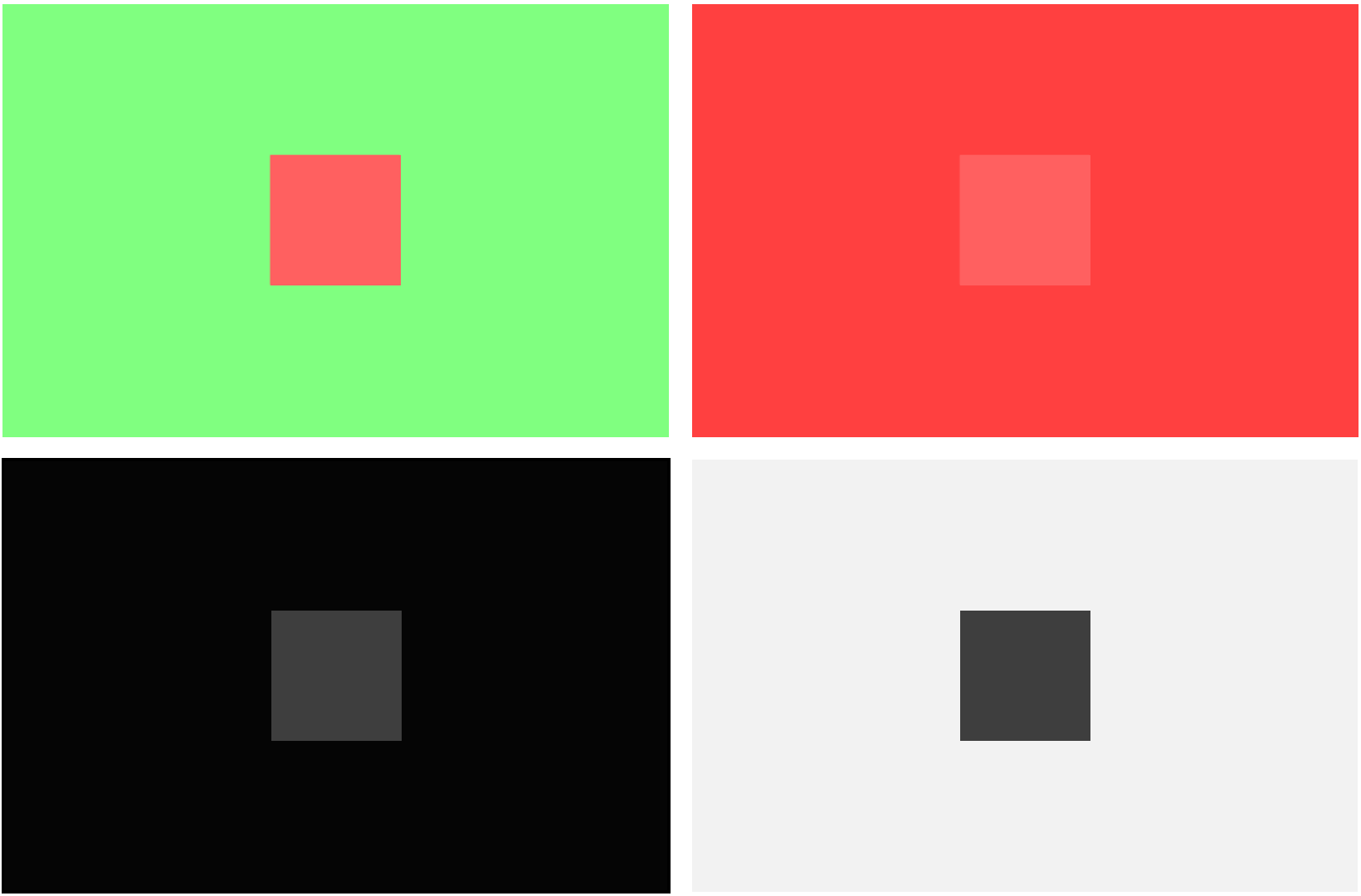}
\caption{Demonstration of the simultaneous color induction. \textbf{Top:} The two red patches in the center are displayed in the same RGB value $(255,96,96)$, yet they appear different colors in perception. \textbf{Bottom:} Two grey patches are displayed in the same RGB value $(64,64,64)$ but have different perceived colors. More patterns can be found in related studies \cite{Brown1997,Webster2002,Ekroll2013,Klauke2015}.}
\label{fig:colorinduction}
\vspace{-0.6cm}
\end{figure}

\textbf{Color Contrast.} Color contrast is the perceptual difference between one region and its adjacent region in color. In the HVS, this difference is influenced by the chromaticity and luminance of a test stimulus and its surrounding area when presented simultaneously \cite{Jameson1959, Krauskopf1986, Jameson1961}. This effect is called simultaneous color induction. For example, a red patch looks redder on a green background than on a red background (see Figure~\ref{fig:colorinduction}). Many studies verify the color induction and related phenomenon through various experiments \cite{Krauskopf1986, Brown1997}. Some researchers systematically quantify the effects of simultaneous color induction along the dimension of hue \cite{Webster2002, Klauke2015}.

\zyj{It is generally accepted that} two contents with different colors displayed simultaneously induce a complementary color shift to each other (\zyj{the complementarity law}). In other words, changes in the color appearance of the induced stimulus are directed away from the appearance of inducing surrounds. \zyj{Recently, some hypotheses, such as the direction law \cite{Ekroll2012,Ekroll2013,Ratnasingam2017}, challenge the traditional complementarity law, showing that the mechanism of simultaneous color induction is still not understood completely. Additionally, verifying these hypotheses is beyond the scope of this work.} Nevertheless, these valuable advances provide a theoretical and experimental foundation for our approach, prompting us to stand on a novel perspective for color blending.


\section{Problem Statement}
\label{sc:problem}

\begin{figure}
  \centering
  \includegraphics[width=\columnwidth]{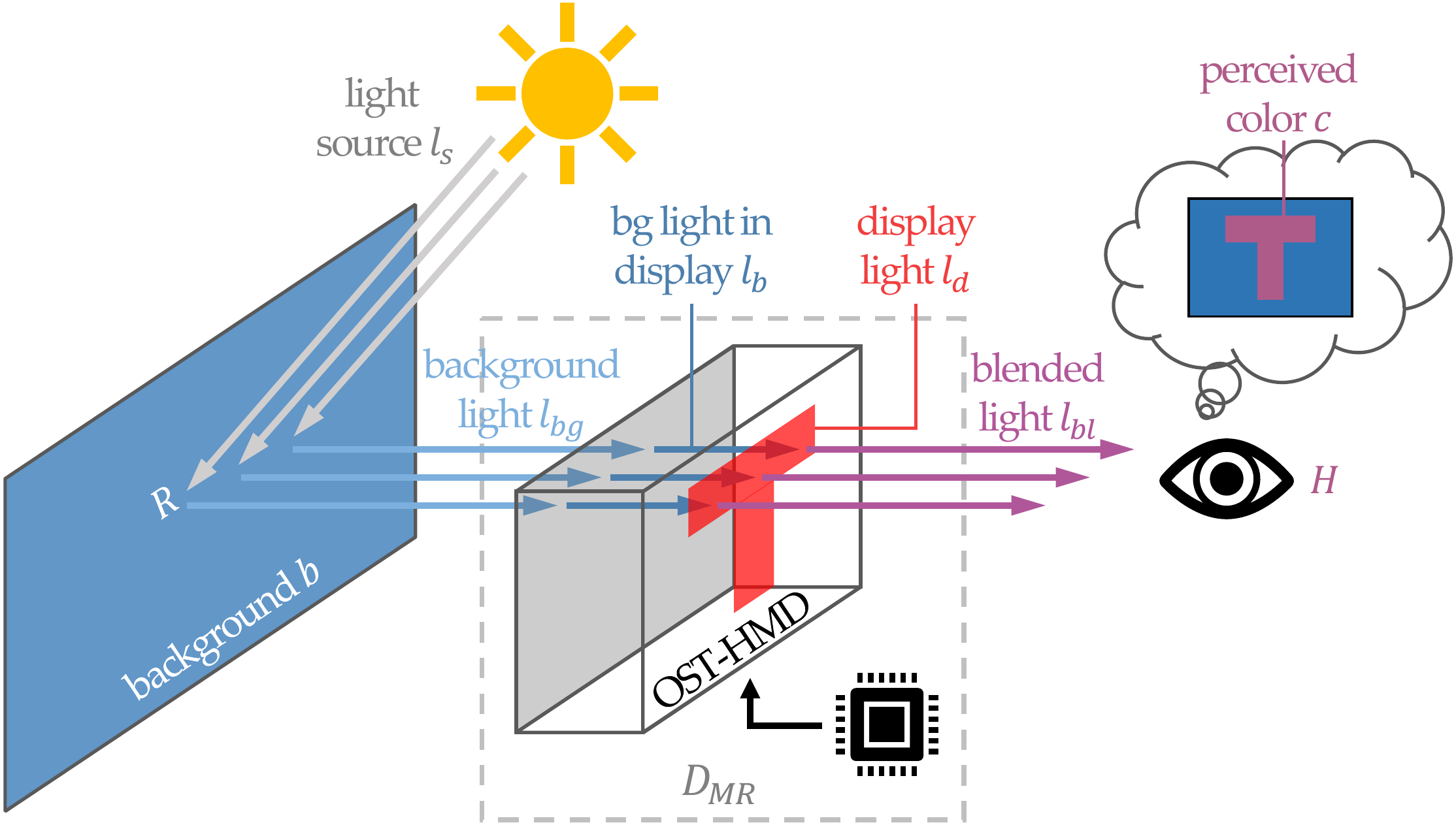}
  \caption{Illustration of the color-blending problem on OST-HMDs. The light source $l_s$ interacts with the object surface in the background $b$ by the reflectance function $R$, and then the reflected light (i.e., the background light $l_{bg}$) enters the OST-HMDs. Subsequently, the entire display system $D_{M\!R}$ blends the background light in displays $l_b$ with the display light $l_d$ to generate the blended light $l_{bl}$. Finally, the user's eyes receive $l_{bl}$, which forms the perceived color $c$ through the operation $H$ of HVS. }
  \label{fig:colorBlending}
  \vspace{-0.6cm}
  \end{figure}

In this section, we describe the color-blending problem illustrated in Figure~\ref{fig:colorBlending}. We formulate our method, color contrast enhancement, as a constrained optimization of color blending. According to the definition of color blending on OST-HMDs by Gabbard et al. \cite{Gabbard2010}, the blending procedure can be formulated as follows:
\begin{subequations}\label{eq:colorblend}
  \begin{align}
  & c = H\left(l_{bl}\right) \\
  & l_{bl} = D_{M\!R}\left(l_{d},l_{bg}\right) \\
  & l_{bg} = R\left(l_{s},b\right) ,
  \end{align}
\end{subequations}
where $c$ represents the 
perceived color, and $H$ is the operation of the HVS for the blended light $l_{bl}$ that reaches the user's eyes. $D_{M\!R}$ is an abstract of the entire OST-HMDs system, which contains multiple parameters, such as lens opacity and display brightness. $l_{d}$ denotes the display light and $l_{bg}$ refers to the background light that enters the front of the OST-HMDs. Reflectance function $R$ depicts how the light source $l_{s}$ in the background $b$ interacts with the object surface and finally enters the OST-HMDs. Previous software-based color correction methods devote to accurately measuring and estimating $l_{bg}$ and parameters of $D_{M\!R}$ (e.g., lens opacity), allowing them to subtract the background light in displays ($l_{b}$) from the display light ($l_{d}$). As a result, these methods remove the background color from the display pixels to mitigate the color blending effect. Unlike color correction approaches, we are not seeking to handle the distortions \cite{Hincapie-Ramos2015} or measure hardware parameters of OST-HMDs \cite{Langlotz2016}. Instead, we focus on the HVS's operation $H$ and the perceived color $c$. Therefore, function $D_{M\!R}\left(l_{d},l_{bg}\right)$ can be approximated as $l_{d}+l_{b}$:
\begin{align}\label{eq:simplify}
  l_{bl} \approx l_{d}+l_{b} .
\end{align}

On the basis of the opponent-colors theory proposed by Jameson and Hurvich \cite{Jameson1959}, the perceived color $c$ of a test stimulus can be expressed as follows:
\begin{align}\label{eq:opponentcolor}
  c=f[ {(r-g)}_t+{(r-g)}_i,{(y-b)}_t+{(y-b)}_i, & \notag \\
  {(w-bk)}_t+{(w-bk)}_i] & ,
\end{align}
where $f$ is a function of the sums. \zyj{${(r-g)}_t$, ${(y-b)}_t$, and ${(w-bk)}_t$ are responses of three paired and opponent neural systems of the test stimulus. ${(r-g)}_i$, ${(y-b)}_i$, and ${(w-bk)}_i$ denote the responses of corresponding systems from the surrounding area induced by the simultaneous color induction. It is vice versa that this induction also affects the perceived color of the surrounding area.}

The color difference $\Delta E^\ast_{ab}$ defined in the CIEL$^\ast$a$^\ast$b$^\ast$ color space (abbreviated as the LAB color space) between color $x$ and color $y$ can be calculated as follows: 
\begin{align}\label{eq:deltaE}
  \Delta E^\ast_{ab}(x, y) = &\; ||x - y|| \\
   = &\; \sqrt{{(L_{x}^{\ast} - L_{y}^{\ast})}^2 + {(a_{x}^{\ast} - a_{y}^{\ast})}^2 + {(b_{x}^{\ast} - b_{y}^{\ast})}^2}, \notag 
\end{align}
where $L^{\ast}$, $a^{\ast}$, and $b^{\ast}$ are three orthonormal bases of the LAB color space used to describe the luminance and chromaticity of colors.

The key of our approach is to \zyj{shift $l_{d}$ for increasing the color difference $\Delta E^\ast_{ab}(l_{d}, l_{b})$ in \eqref{eq:deltaE}. In addition, shifting $l_{d}$ leads to an increase in the corresponding responses ${(r-g)}_i$, ${(y-b)}_i$, and ${(w-bk)}_i$ in \eqref{eq:opponentcolor}, which further enhances the perceived color difference between $l_{d}$ and surrounding $l_{b}$. In this manner, we improve the distinction between the virtual content and surrounding background. We've also tested enhancing $\Delta E^\ast_{ab}(l_{d}, l_{d}+l_{b})$ and find that $l_{d}+l_{b}$ usually has a large luminance component but a relatively small chromaticity component. Applying our color contrast enhancement for such a color usually produces brighter colors than that produced from $l_d$, but decreases the contrast within surfaces of virtual objects. Therefore, we choose to enhance $\Delta E^\ast_{ab}(l_{d}, l_{b})$ instead of $\Delta E^\ast_{ab}(l_{d}, l_{d}+l_{b})$.}

\begin{figure}
  \centering 
  \includegraphics[width=\columnwidth]{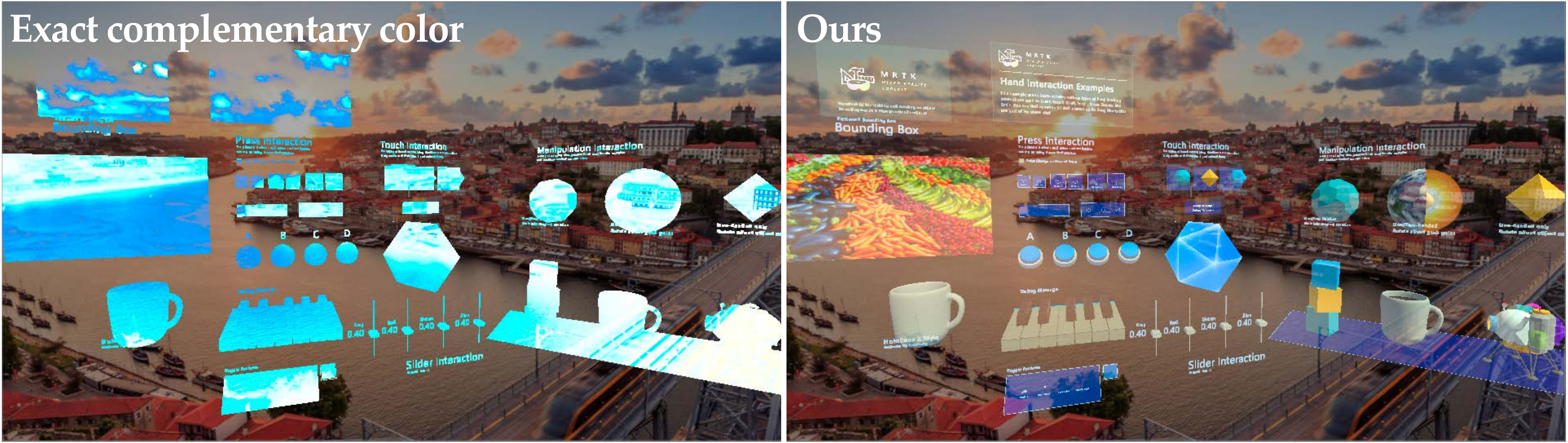}
  \caption{Comparison between using the exact complementary color and our optimization. We use a landscape photo as the test background and some objects from the \textit{Hand Interaction Examples}\protect\footnotemark~scene as the virtual content. The rendering results in the left uses the per-pixel complementary color of the background color. The image shown in the right is rendered with our method. Please refer to the supplementary video for an example in the real environment captured by the HoloLens.}
  \label{fig:x_o}
  \vspace{-0.6cm}
\end{figure}

When the background light in displays $l_{b}$ and the display light $l_{d}$ exhibit the most significant color difference, that is, $l_{b}$ and $l_{d}$ are complementary colors, the HVS perceives a maximum color contrast. However, considering only the color contrast may lead to an unintended color altering of displayed objects. Figure~\ref{fig:x_o} shows an example of such a case. It is clear that the virtual objects at the bottom left lack texture details and can hardly be recognized. This most straightforward optimization rule can almost only be used for textual content. Therefore, we introduce several constraints in the optimization of optimal displayed color $l_{opt}$ to maintain the color consistency between the enhanced color and the original color as:
\begin{align}\label{eq:optimization}
  l_{opt} & = \argmax_{l_{opt}} \Delta E^\ast_{ab}(l_{opt}, l_{b}) && \text{subject to constraints} .
\end{align}

\footnotetext{https://github.com/microsoft/MixedRealityToolkit-Unity}
%

\subsection{Constraints of the Optimization}
\label{ssc:constraints}

Given the aforementioned issue, the color difference $\Delta E^\ast_{ab}$ between $l_{b}$ and $l_{d}$ cannot be unlimited. One should keep the color consistency of the original color of virtual objects. On this basis, we introduce the first constraint, named the \textbf{Color Difference Constraint}, to restrict the range of $l_{opt}$:
\begin{align}\label{eq:color_difference_constraint}
   \Delta E^\ast_{ab}(l_{opt},l_{d})\le\lambda_E ,
\end{align}
where $\lambda_E$ represents a non-negative color difference threshold. As such, the optimal displayed color $l_{opt}$ and the original displayed color $l_{d}$ should be kept within a certain range in color.

\begin{figure}
  \centering
  \includegraphics[width=\columnwidth]{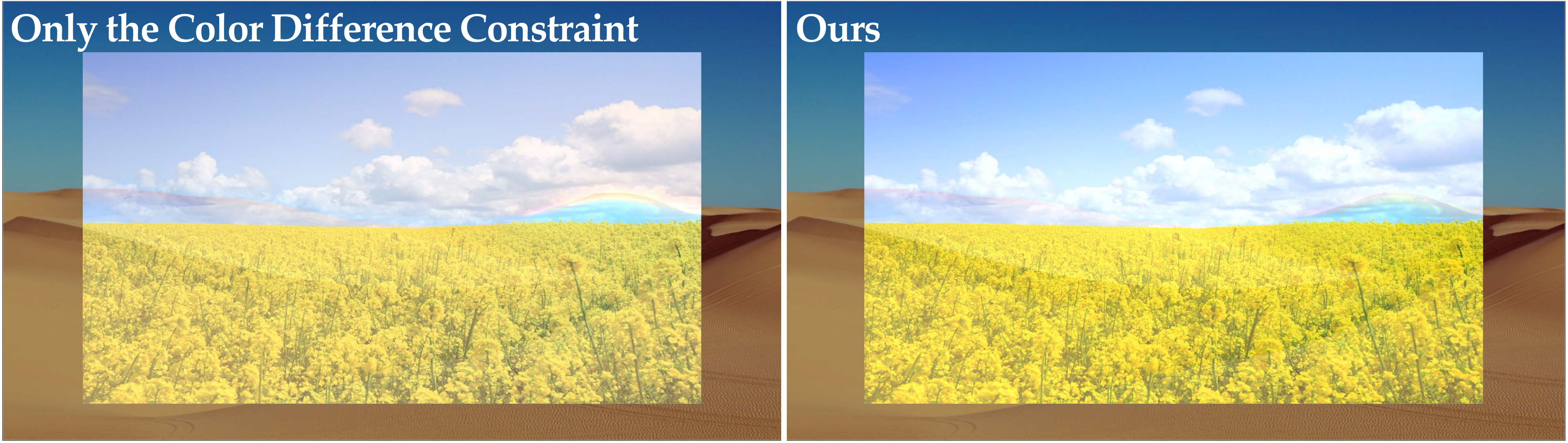}
  \caption{Demonstration of the Chroma Constraint. Two landscape photos serve as the test background and virtual content. The left side is rendered with the optimal colors that meet only the Color Difference Constraint, and the right side is the optimized result of our method. Please refer to the supplementary video for an example in the real environment captured by the HoloLens.}
  \label{fig:patch3}
  \vspace{-0.6cm}
\end{figure}

Further, if the hues of the background color $l_{b}$ and the display color $l_{d}$ are similar, the optimized $l_{opt}$ would shift along the complementary direction of $l_{b}$, resulting in a decrease in chroma of $l_{d}$ (see Figure~\ref{fig:patch3} for a specific example). We introduce the second constraint to tackle the chroma reduction, named the \textbf{Chroma Constraint}: 
\begin{align}\label{eq:chroma_constraint}
  {ch}_{opt}-{ch}_{d}\geq 0 ,
\end{align}
where ${ch}_{opt}$ and ${ch}_{d}$ represent the chroma of $l_{opt}$ and $l_{d}$, respectively. 
Note that although we have restricted the reduction in chroma, there is no boundary to the increment existing with this constraint.

\begin{figure}
  \centering
  \includegraphics[width=\columnwidth]{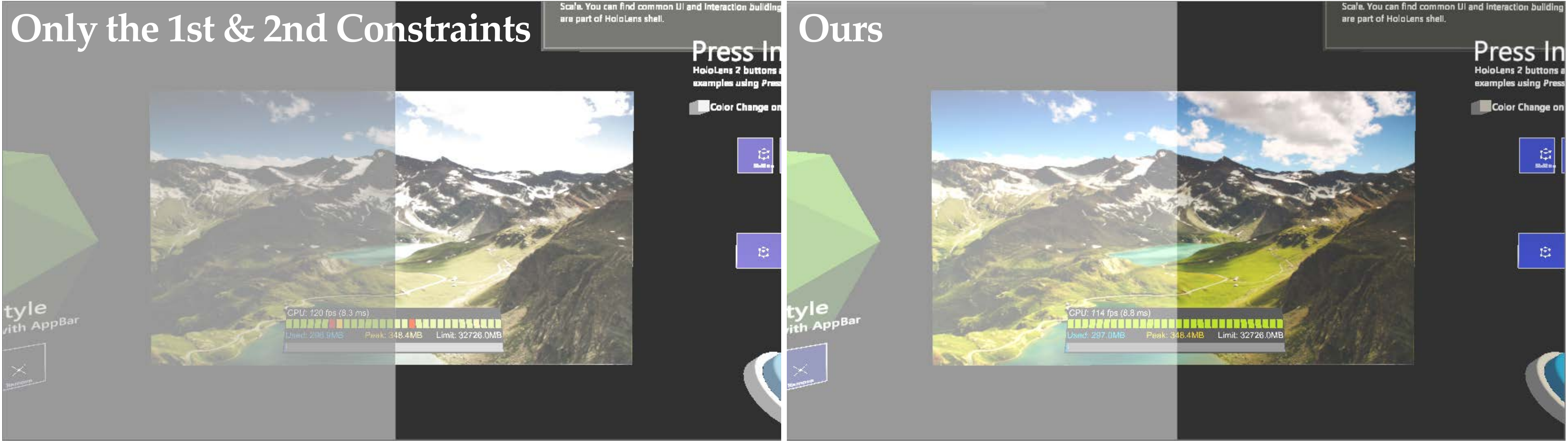}
  \caption{Demonstration of the Luminance Constraint. We use a bright gray and a dark gray as the test background and a landscape photo and other objects as the virtual contents. The left side is rendered with the optimal colors that follow the first and second constraints but not the third, and the right side is the optimized result of our algorithm. Please refer to the supplementary video for an example in the real environment captured by the HoloLens.}
  \label{fig:patch1}
  \vspace{-0.6cm}
\end{figure}

However, in addition to the constraint added on chroma, the luminance can benefit from adding a corresponding constraint. For example, if $l_{b}$ is a bright, whitish color, such as a wall or clear sky, the complementary color of $l_{b}$ is close to a dark grey. In this case, the optimized color $l_{opt}$ moves toward the direction of $l_{b}$'s complementary color, resulting in a decrease in luminance (see Figure~\ref{fig:patch1} for an example). Reducing the luminance of displayed color on common OST-HMDs often leads to increased transparency. Therefore, displaying $l_{opt}$ with no constraint on luminance reduces the visibility of virtual objects. Moreover, the experimental results of Fukiage et al. \cite{Fukiage2014} showed that if the visibility of virtual objects in OST-HMDs is enhanced by unconstrainedly increasing the luminance of $l_{d}$, then the perceptual contrast within surfaces of these objects decreases. These situations lead to the third constraint, which we call the \textbf{Luminance Constraint}:
\begin{align}\label{eq:luminance_constraint}
  \Delta L^\ast(l_{opt},l_{d})\le\lambda_{L},
\end{align}
where $\Delta L^\ast$ is the luminance difference, and $\lambda_{L}$ denotes a non-negative luminance threshold. This constraint alleviates reduced visibility of the displayed content caused by the bright background, as well as the contrast reduction of virtual objects in a dim environment. 

Finally, we introduce an evident constraint between $l_{opt}$ and $l_{b}$, named the \textbf{Just Noticeable Difference Constraint}:
\begin{align}\label{eq:jnd_constraint}
  \Delta E^\ast_{ab}(l_{opt},l_{b})\geq\lambda_{J\!N\!D} ,
\end{align}
where $\lambda_{J\!N\!D}$ represents the just noticeable difference (JND). This constraint indicates that the optimal color $l_{opt}$ and the background color $l_{b}$ require a minimum color difference that the HVS can distinguish. Note that this constraint is generally satisfied automatically on account of the objective of our optimization. However, for certain extreme cases where users exhibiting an above-average JND, applying this constraint is mandatory.

\section{Algorithm}
\label{sc:algorithm}
%

\begin{figure*}
  \centering
  \includegraphics[width=\textwidth]{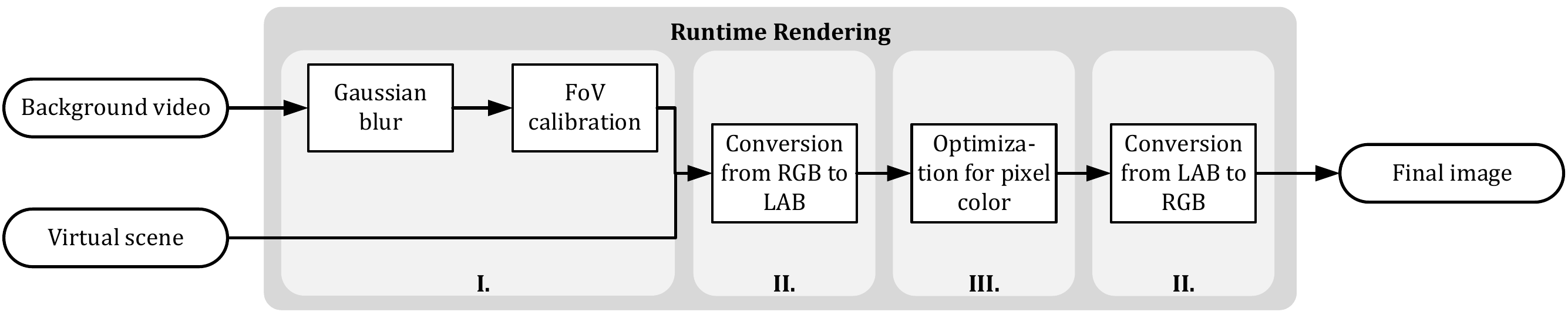}
  \caption{Algorithm Overview. Our algorithm takes the rendered virtual scene and the streaming background video as input. First, a Gaussian blur and FoV calibration are applied to the background video. Second, we transform the blurred video and the virtual scene from RGB to the LAB color space and then optimize the displayed color for all pixels. Finally, the pixel color of the virtual scene is converted back to the RGB color space and output to displays.}
  \label{fig:overview}
  \vspace{-0.4cm}
\end{figure*}

Given these four constraints of color contrast enhancement, the next step is to solve the optimization problem, as shown in~\eqref{eq:optimization}. Accordingly, we design a real-time algorithm to enable color contrast enhancement in a variety of real environments. 
Figure~\ref{fig:overview} shows an overview of the proposed method, which mainly includes three procedures:

\begin{enumerate}[label=\textbf{\Roman*.}, align=left, leftmargin=*]
    \item \textbf{Preprocessing:} We perform a Gaussian blur and field of view (FoV, see Subsection~\ref{ssc:Preprocessing}) calibration for the streaming video of the background.
    \item \textbf{Conversion:} We convert the display and background colors from RGB to the LAB color space and vice verse.
    \item \textbf{Optimization:} Utilizing the aforementioned four constraints, we optimize the displayed colors on the basis of the background colors.
\end{enumerate}

\subsection{Preprocessing}
\label{ssc:Preprocessing}


\begin{figure}[tb]
  \centering 
  \includegraphics[width=\columnwidth]{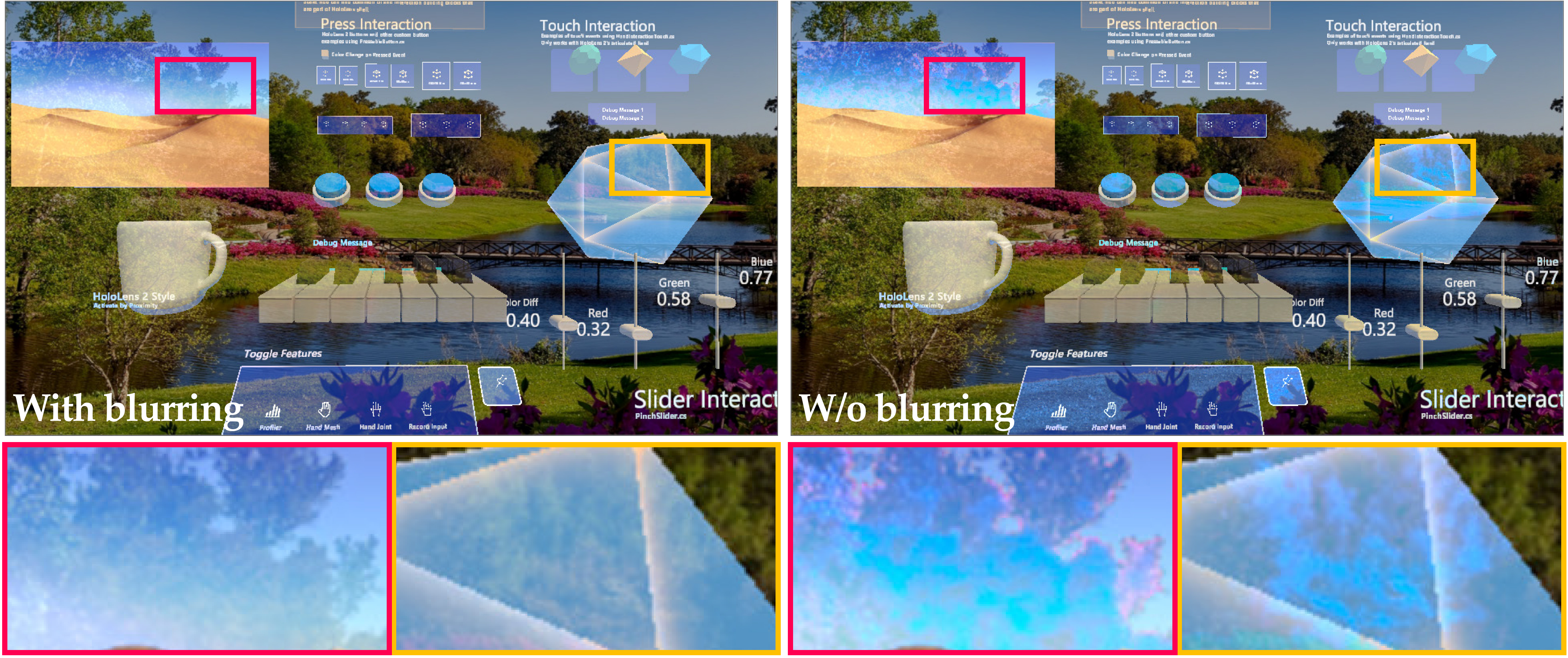}
  \caption{Demonstration of the necessity of image blur. We use a landscape photo as the test background. The left side is the optimized result of our algorithm with background blurring enabled. Optimal displayed colors without background blurring are represented on the right side. The zoomed regions emphasize details.}
  \label{fig:GaussianBlur}
  \vspace{-0.2cm}
\end{figure}

It is generally believed that multiple individual pattern analyzers contribute to the contrast sensitivity of humans \cite{Campbell1968}. These pattern analyzers are often called spatial-frequency channels, which filter the perceived image into spatially localized receptive fields with a limited range of spatial frequencies. That is, different analyzers are tuned to various types of information. For example, the low-spatial-frequency channels receive the color and outlines of objects, whereas the high-spatial-frequency channels perceive details. Considering the low-pass nature of color vision \cite{Anderson1991} \zyj{and the blurring characteristic of the non-focal field in the HVS, our method does not need a pixel-precise camera-to-display calibration.} Instead, we use a Gaussian blur to \zyj{extract the spatial color information and} filter out details of the background video as 
$G(x,y)=\frac{1}{2\pi\sigma^2}e^{-\frac{x^2+y^2}{2\sigma^2}}$, 
where $x$ and $y$ are the horizontal and vertical distances from a pixel to its center pixel, respectively, and $\sigma$ is the standard deviation of the normal distribution. \zyj{Such a blurring simulates the non-focal effect of the HVS.} 
In this manner, the displayed color obtains the weighted average of multiple background colors in the corresponding region. This filter also reduces the flicker in optimized colors caused by the high-spatial-frequency details in the background (see Figure~\ref{fig:GaussianBlur} for examples).

\begin{figure}[tb]
  \centering 
  \includegraphics[width=\columnwidth]{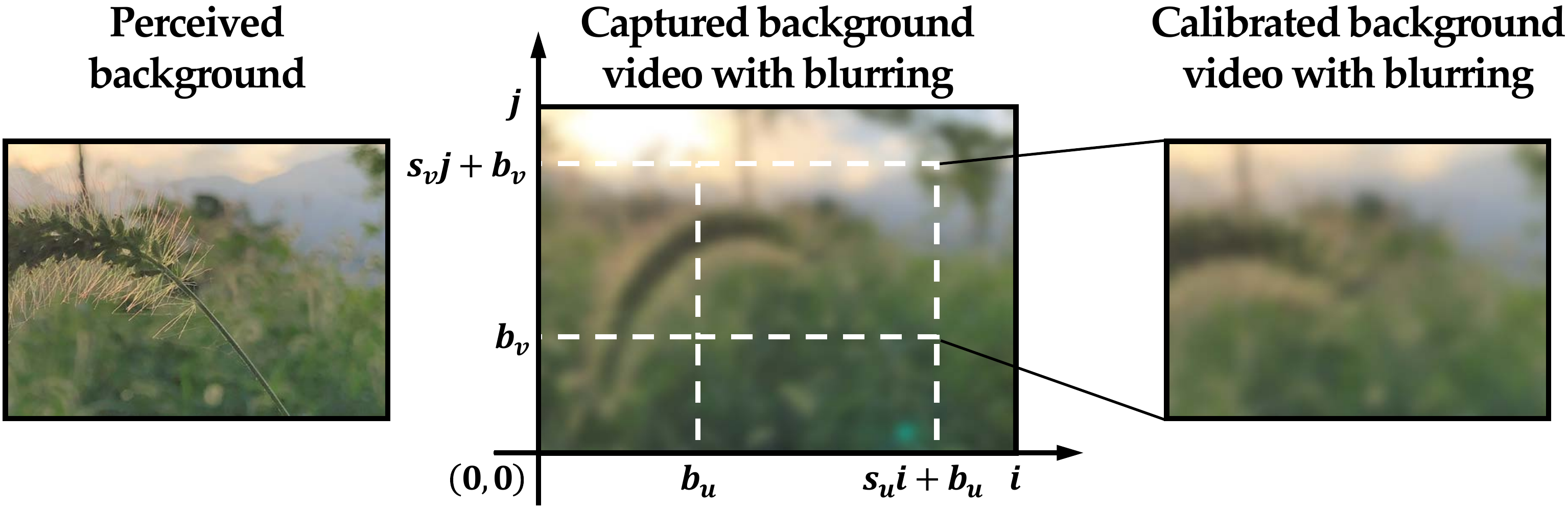}
  \caption{Illustration of FoV calibration. This process is performed when the FoV between the captured background video and the frame buffer are different. We crop and scale the captured video to match the position and size of the background seen through OST-HMDs.}
  \label{fig:fovCalibration}
  \vspace{-0.6cm}
\end{figure}

Given the low-frequency characteristics of the blurred background video and the difference of FoV between captured videos and OST-HMDs, we subsequently use a screen-space coordinate mapping called FoV calibration between the background video and the frame buffer of rendering system, in order to approximate pixel-precise calibration (such as homography). On this basis, we apply the following coordinate mapping to map the background video to the frame buffer:
\begin{align}\label{eq:calibration}
  \begin{cases}
      u = s_u i + b_u  \\
      v = s_v j + b_v,
  \end{cases}
\end{align}
where $(u,v)$ and $(i,j)$ represent the 2D texture coordinates of the frame buffer and the background video, respectively. $(s_u,s_v)$ denotes the scale factor in the horizontal and vertical directions of the background video coordinates, and $(b_u,b_v)$ is corresponding offsets. Figure~\ref{fig:fovCalibration} shows the illustration. We assume that the FoV of captured videos is greater than that of OST-HMDs. Using this calibration, the low-frequency information of background videos and that of real scenes seen through displays of OST-HMDs is as similar as possible.

\subsection{Conversion}
\label{ssc:Conversion}

After preprocessing, the displayed color can be paired with an average background color of the corresponding area. These colors are all stored and represented in the RGB color space. However, the widely used RGB color space is not perceptually uniform that the same amount of numerical change does not correspond to the same amount of color difference in visual perception. 

In order to achieve better optimization, we convert the background color and the displayed color from RGB to the LAB color space, to take full advantage of its perceptual uniformity and independence of luminance and chromaticity. After performing the color contrast enhancement, we transform LAB colors back to RGB to display appropriately on OST-HMDs.

Additionally, the gamut of the LAB color space is more extensive than displays and even the HVS, indicating that many of the coordinates in the LAB color space, \zyj{especially those located in the edge area,} cannot be reproduced on typical displays. For simplicity, we scale the range of the original LAB color space to $\left[-1,1\right]$ and take the inscribed sphere as the solution space of our method. 

\subsection{Optimization}
\label{ssc:Optimization}


\begin{figure*}[tb]
  \centering

  \subfloat[]{
  \includegraphics[width=0.49\columnwidth]{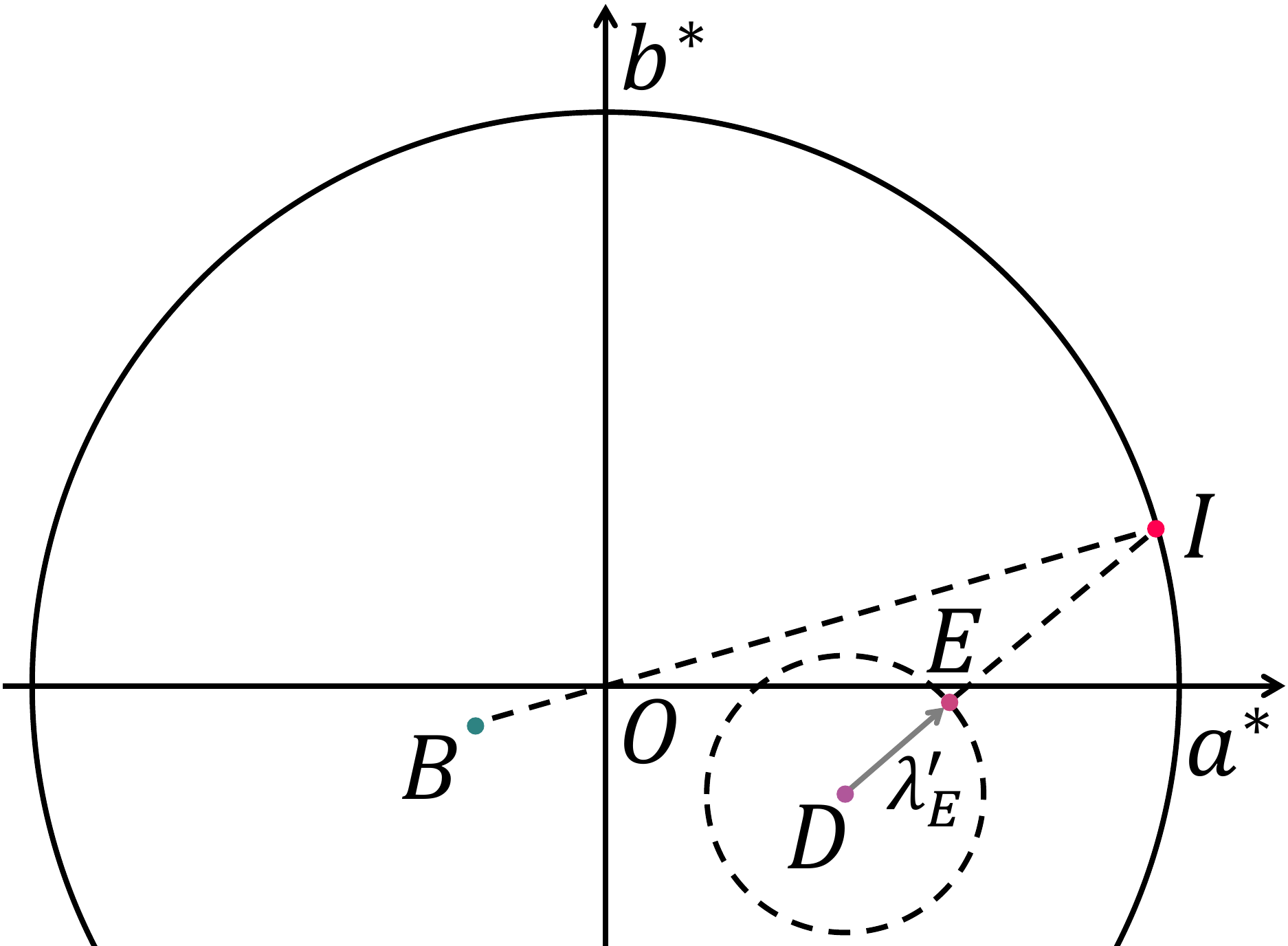}
  \label{fig:LiLc}
  }
  \subfloat[]{
  \includegraphics[width=0.49\columnwidth]{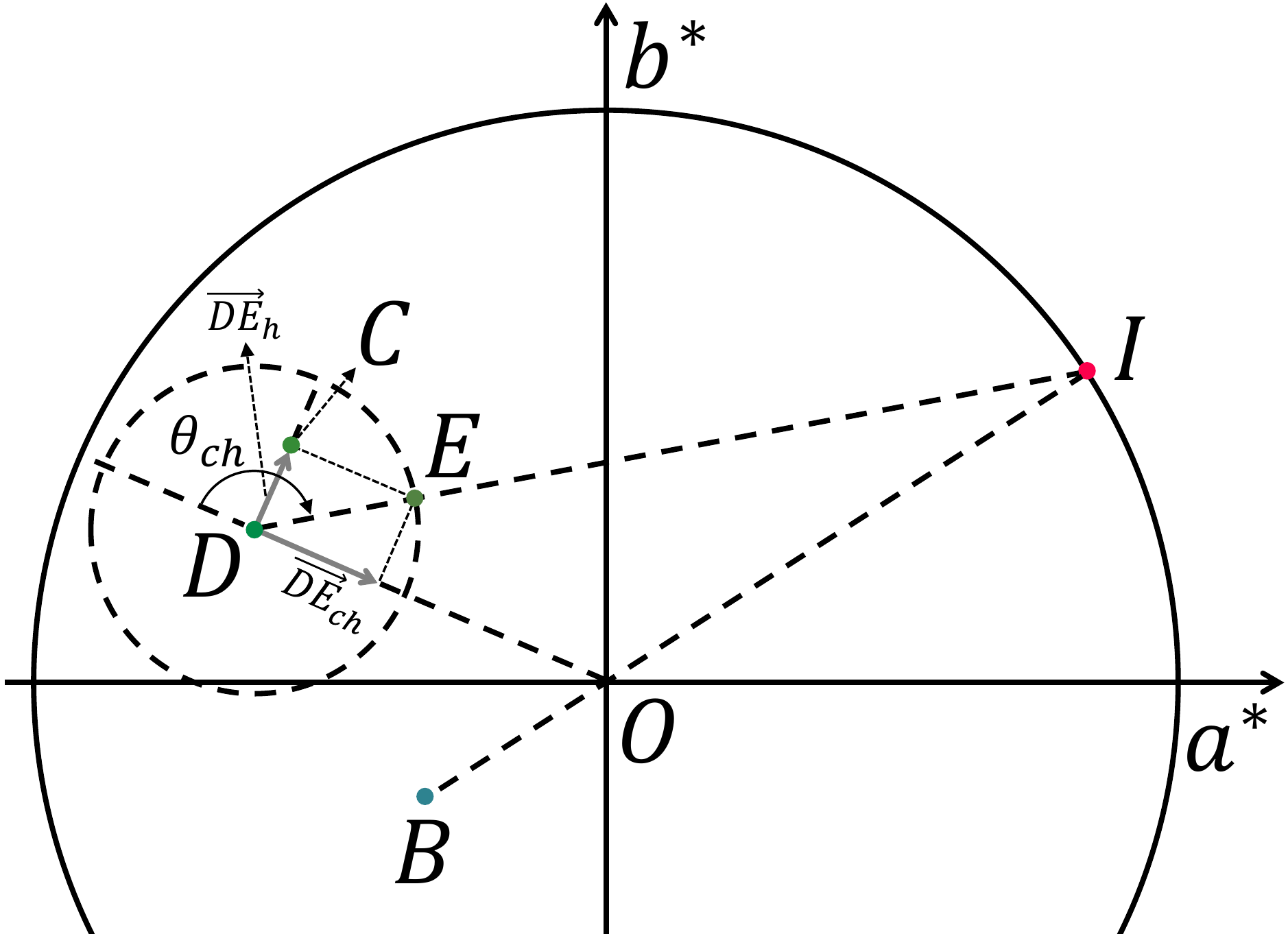}
  \label{fig:Sab}
  }
  \subfloat[]{
  \includegraphics[width=0.49\columnwidth]{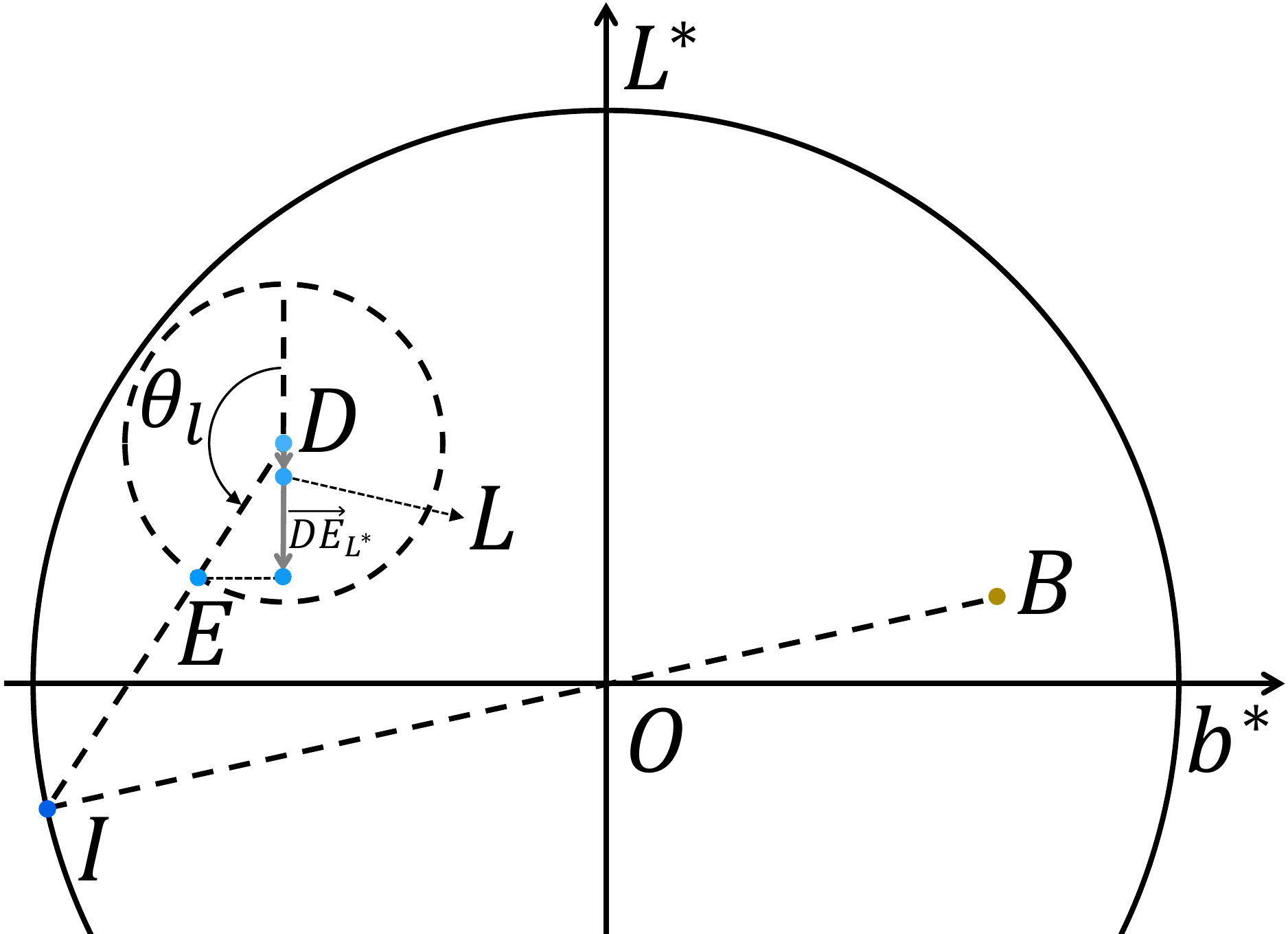}
  \label{fig:Bl}
  }
  \subfloat[]{
  \includegraphics[width=0.49\columnwidth]{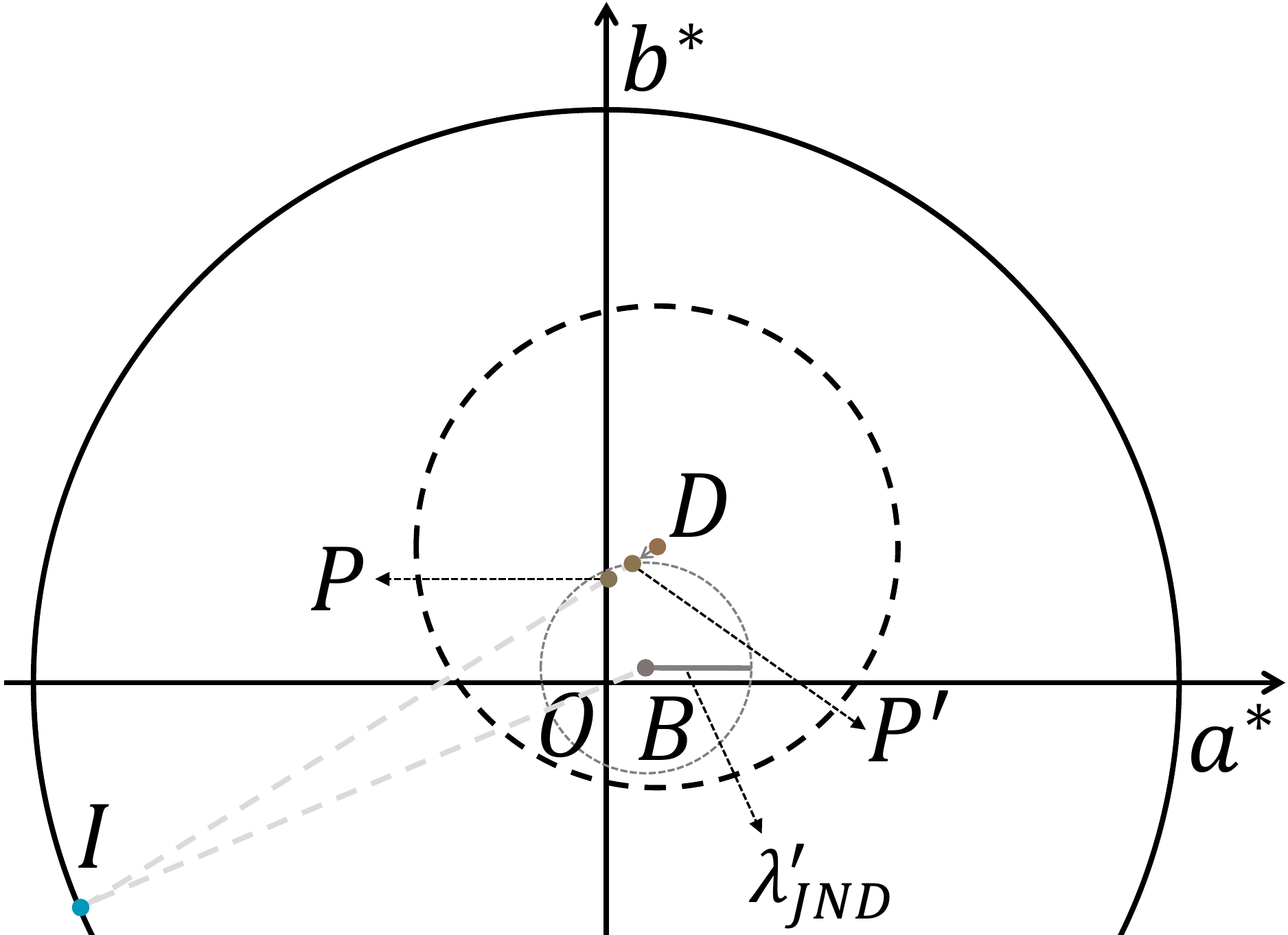}
  \label{fig:JND}
  }
  \caption{
    We illustrate our algorithm on the 2D plane formed by two of the three axes of the LAB color space for simplicity. 
    (\textbf{a}) Illustration of calculating coordinates $I$ and $E$. All points in this subfigure are plotted on the plane $a^\ast O b^\ast$. In the unit circle, $I$ is the point farthest from $B$. All points with the same distance $\lambda^\prime_E$ from $D$ form a circle, which intersects the line $DI$ at point $E$. In this figure, all points are colored with the same $L^\ast$ value equals to neutral gray. 
    (\textbf{b}) Illustration of determining the coordinates $C$. All points in this subfigure are plotted on the plane $a^\ast O b^\ast$. If $\theta_{ch}$ is an obtuse angle, the change in chroma $\protect\overrightarrow{DE}_{ch}$ of vector $\protect\overrightarrow{DE}$ will be discarded. Thus, the chroma-constrained vector $\protect\overrightarrow{DC}$ only contains the component $\protect\overrightarrow{DE}_{h}$ of $\protect\overrightarrow{DE}$. In this case, all points are colored with the same $L^\ast$ value equals to neutral gray. 
    (\textbf{c}) Illustration of determining the coordinates $L$. All points in this subfigure are plotted on the plane $b^\ast O L^\ast$. The change in luminance $\protect\overrightarrow{DE}_{L^\ast}$ of vector $\protect\overrightarrow{DE}$ will be attenuated with $\cos{\theta_l}$. Thus, $\protect\overrightarrow{DE}_{L^\ast}$ is shortened to $\protect\overrightarrow{DL}$. In this case, all points are colored with the same $a^\ast$ value equals to zero. 
    (\textbf{d}) Illustration of finding the new optimal  displayed color $P^\prime$ under the Just Noticeable Difference Constraint. All points in this subfigure are plotted on the plane $a^\ast O b^\ast$. Sometimes the coordinates $P$ of optimal displayed color $l_{opt}$ falls into the circle $B$ on account of a large radius $\lambda^\prime_{J\!N\!D}$, meaning that the HVS cannot distinguish between $l_{opt}$ and $l_b$. Therefore, we find the intersection of line $DP$ and circle $B$ as $P^\prime$. In this figure, all points are colored with the same $L^\ast$ value equals to neutral gray.
  }
  \vspace{-0.6cm}
\end{figure*}


Given a blurred background color $l_b$ and an original displayed color $l_d$ in the scaled LAB color space, the objective of our optimization is to find the optimal displayed color $l_{opt}$. 

In this work, we denote points by capital italic letters. First, we calculate the coordinates $I$ of the ideally optimal displayed color corresponding to the blurred background color without considering any constraint:
\begin{subequations}
  \begin{align}\label{eq:ideal_L}
    I &=-\frac{B}{\text{dist}(B,O)}  \\
    &= -\text{norm}(\overrightarrow{OB}),
  \end{align}  
\end{subequations}
where $B$ represents the coordinates of the blurred background color $l_{b}$. $\text{dist}(B,O)$ denotes the distance between $B$ and the center $O$ of the unit sphere. This formula can also be rewritten to the latter form, where $\text{norm}(\overrightarrow{OB})$ means to normalize the vector $\overrightarrow{OB}$. That is, the farthest point from $B$ in the unit sphere is the intersection of the extension line of $BO$ and the sphere (Figure~\ref{fig:LiLc}).


Given the ideally optimal displayed color $I$, we then consider the aforementioned four constraints: Color Difference Constraint, Chroma Constraint, Luminance Constraint, and Just Noticeable Difference Constraint, into color optimization in~\eqref{eq:optimization}. First, we calculate the coordinates $E$ of initially optimal display color by applying the Color Difference Constraint to $I$:
\begin{align}\label{eq:L^c}
  \overrightarrow{DE}=\min{(\text{dist}(D,I),\lambda^\prime_E)}\cdot \text{norm}(\overrightarrow{DI}),
\end{align}
where $\overrightarrow{DI}$ is the ideal change vector starting from the coordinates $D$ of original displayed color $l_{d}$, and $\lambda^\prime_E$ is the scaled color difference threshold specified by users. Now, we have the change vector $\overrightarrow{DE}$ of the displayed color constrained by the color difference. Figure~\ref{fig:LiLc} presents a two-dimensional illustration of this step. 

Let $\textbf{v}^{\prime}$ be the projection of vector $\textbf{v}$ onto the plane $a^\ast O b^\ast$ of the LAB color space. Thus, the change in chroma $\overrightarrow{DE}^{\prime}_{ch}$ of the vector $\overrightarrow{DE}^{\prime}$ can be obtained by calculating the projection of $\overrightarrow{DE}^{\prime}$ onto the vector $\overrightarrow{OD}^{\prime}$ (see Figure~\ref{fig:Sab} for a two-dimensional example). For the Chroma Constraint, we discard the chroma reduction of $\overrightarrow{DE}^{\prime}$ as:
\begin{subequations}
  \begin{align}\label{eq:Sab}
    & \overrightarrow{DC}=t_{ch}\cdot \overrightarrow{DE}^{\prime}_{ch} + \overrightarrow{DE}^{\prime}_{h} \\
    & t_{ch} = 
    \begin{cases}
      1,  & \mbox{if }\theta_{ch}\leq 90^{\circ} \\
      0,  & \mbox{if }\theta_{ch} > 90^{\circ}.
    \end{cases}
  \end{align}
\end{subequations}

\begin{algorithm}[H]
  \caption{Finding the coordinates $P$ of the optimal displayed color $l_{opt}$}
  \label{alg:optimization}
  \begin{algorithmic}[1]
  \REQUIRE $D, B, \lambda^\prime_E, \lambda^\prime_{J\!N\!D}$
  \ENSURE  $P$
   \\ \textit{Components $x$, $y$, and $z$ denote $L^\ast$, $a^\ast$, and $b^\ast$, respectively.}
   \STATE $dir = $ Vector3(0, 0, 0);
   \\ \textit{Color Difference Constraint} :
   \STATE $I = -$ norm($B$);
   \STATE $E = $ min$(\text{dist}(D,I),\lambda^\prime_E)$ $\cdot$ norm($I - D$);
   \\ \textit{Chroma Constraint} :
   \STATE $E^{\prime} = $ Vector3(0, $E.yz$);
   \STATE $E^{\prime}_{ch} = $ change in chroma of $E^{\prime}$;
   \STATE $E^{\prime}_{h} = E^{\prime} - E^{\prime}_{ch}$;
   \STATE $\theta_{ch} = $ angle from Vector3(0, $D.yz$) to $E^{\prime}$;
   \STATE $dir$.$yz = $ (($\cos{\theta_{ch}} \geq$ 0 ? 1 : 0) $\cdot$ $E^{\prime}_{ch}$ + $E^{\prime}_{h}$)$.yz$;
   \\ \textit{Luminance Constraint} :
   \STATE $\theta_{l} = $ angle from Vector3(1, 0, 0) to $E$;
   \STATE $dir.x = $ ($\cos{\theta_{l}} \geq$ 0 ? (1 - $\cos{\theta_{l}}$) : (1 + $\cos{\theta_{l}}$)) $\cdot$ $E.x$;
   \\ \textit{Optimal display color} :
   \STATE $P = D + dir$;
   \\ \textit{Just Noticeable Difference Constraint} :
   \IF {(dist($P$, $B$) $< \lambda^\prime_{J\!N\!D}$)}
   \STATE $P = $ intersection of line $D\!P$ and circle $B$ of radius $\lambda^\prime_{J\!N\!D}$;
   \ENDIF
  \RETURN $P$
  \end{algorithmic}
\end{algorithm}

Here $\overrightarrow{DE}^{\prime}_{h} = \overrightarrow{DE}^{\prime} - \overrightarrow{DE}^{\prime}_{ch}$, which refers to the component of $\overrightarrow{DE}^{\prime}$ perpendicular to $\overrightarrow{DE}^{\prime}_{ch}$. $\theta_{ch}$ represents the angle ($0^{\circ}$–$180^{\circ}$) from $\overrightarrow{OD}^{\prime}$ to vector $\overrightarrow{DI}^{\prime}$. This angle describes the deviation in hue and chroma between the optimal displayed color and the original displayed color. In this manner, the adaptive parameter $t_{ch}$ 
provides a smooth visual effect when hue and chroma change. 

As for the Luminance Constraint, we also use an adaptive parameter to scale the alterations in luminance:
\begin{align}\label{eq:L_l}
  \overrightarrow{DL}=\left(1-\left|\cos{\theta_l}\right|\right)\cdot \overrightarrow{DE}_{L^\ast},
\end{align}
where $\overrightarrow{DE}_{L^\ast}$ refers to $\overrightarrow{DE}$'s component on the $L^\ast$ axis of the LAB color space. $\theta_l$ represents the angle ($0^{\circ}$–$180^{\circ}$) from the positive $L^\ast$ axis to the vector $\overrightarrow{DE}$ (see Figure~\ref{fig:Bl} for a two-dimensional example). Correspondingly, this angle indicates the difference in luminance between the optimal displayed color and the original displayed color. Once again, this adaptive coefficient $(1-\left|\cos{\theta_l}\right|)$ smoothes the results and makes it redundant to specify the luminance threshold $\lambda_L$ by users.



According to the above steps, the coordinates $P$ of optimal displayed color $l_{opt}$ can be obtained by:
\begin{align}\label{eq:L_opt}
  P=D+\overrightarrow{DC}+\overrightarrow{DL}.
\end{align}

In contrast to the other three constraints, the Just Noticeable Difference Constraint is only applied to some extreme cases (Figure~\ref{fig:JND}), wherein we reduce the length of vector $\overrightarrow{DP}$ so that the distance from the coordinates $P^\prime$ of new optimal displayed color to the coordinates $B$ of the blurred background color is equal to a scaled JND $\lambda^\prime_{J\!N\!D}$ in the unit sphere. Such extreme cases are found in users who have an above-average JND, leading to a larger radius (i.e., $\lambda^\prime_{J\!N\!D}$) of circle $B$ in Figure~\ref{fig:JND}. Algorithm~\ref{alg:optimization} gives the pseudocode of the entire optimization process above, \zyj{where Vector3(0, $E.yz$) generates a three-dimensional vector whose first component is $0$, and the last two components are $E.y$ and $E.z$.}


\section{Implementation}
\label{sc:implementation}

We implemented our algorithm as a full-screen post-processing effect using the Unity 2019$\footnote{https://unity.com/}$ rendering engine and validated it on a software-simulated environment and a commercially available OST-HMD. The simulated environment is based on the Unity Editor with a resolution of $960 \times 540$, using a series of still images as the simulated backgrounds. We used the color difference $\Delta E^\ast_{ab}$ in~\eqref{eq:deltaE} for objective evaluation in the simulated environment. For the real environment, we used the first generation of Microsoft HoloLens MR headset \cite{Kress2017} to evaluate the performance and quality of our algorithm. Both environments take the modified \textit{Hand Interaction Examples} scene as the virtual content, which includes text, photographs, user interfaces, and 3D models with plain or intricate materials. We attenuated the luminance of background images in the simulated environment and streaming videos in the HoloLens by $60\%$ to simulate the real background perceived through the translucent lens in \zyj{the HoloLens. It is a tweakable parameter for other types of OST-HMDs with different transparency.} \zyj{In both environments, the kernel size and $\sigma$ of our Gaussian filter are $3$ and $1.5$, respectively. These values are used for all participants in our user studies (Subsection \ref{ssc:user_study_1} and \ref{ssc:user_study_2}).} We captured streaming videos at $24$ 
frames per second (FPS) by a built-in RGB camera located in the center of the HoloLens in real-time. The screen brightness of the HoloLens was fixed at $70\%$.

\zyj{The data of our FoV calibration (Subsection~\ref{ssc:Preprocessing}) was determined by manual calibration. Specifically, $(s_u,s_v) = (0.65,0.65)$, and $(b_u,b_v) = (0.13,0.17)$.}

\begin{figure*}
  \centering 
  \includegraphics[width=\textwidth]{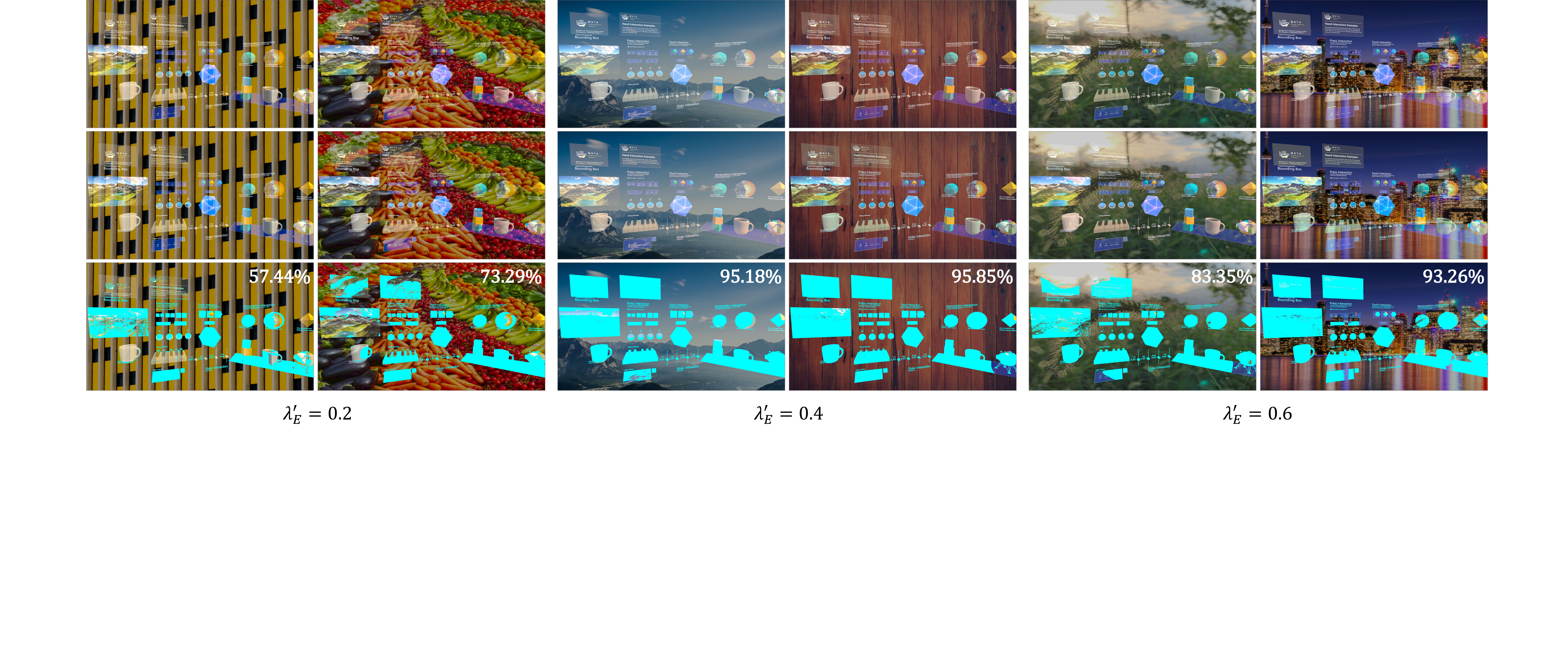}
  \caption{Some of the result images in the simulated environment, generated by the Unity Editor. Each group with a different $\lambda^\prime_E$ shows two background images. The first row shows the original blending images, and the second row shows the results of our method. In the last row, pixels colored in cyan indicate a perceptually increased $\Delta E^\ast_{ab}$ between the background color and the displayed color, wherein numbers in figures represent the corresponding percentage of these pixels in all foreground pixels. Please refer to the supplementary video for an example with different background colors captured by the HoloLens.}
  \label{fig:Results}
  \vspace{-0.6cm}
\end{figure*}

\section{Results}
\label{sc:results}

In our context, we take the scaled color difference ($\lambda^\prime_E$) to regulate our color contrast enhancement, where $\lambda^\prime_E = 0$ means no effect of our algorithm. To demonstrate the effectiveness of our method, we performed a series of experiments in simulated and real environments, including objective evaluations, algorithm comparisons, performance analysis, and subjective user studies. \zyj{All experiments on the HoloLens were conducted in indoor scenes. In real environments, we found that physical environmental conditions (such as scene illuminance) have no significant impact on our experimental results, on account of the auto exposure and the auto white balance of the built-in RGB camera of the HoloLens.}

\subsection{Evaluation}
\label{ssc:evaluation}

\begin{figure}
  \centering 
  \includegraphics[width=\columnwidth]{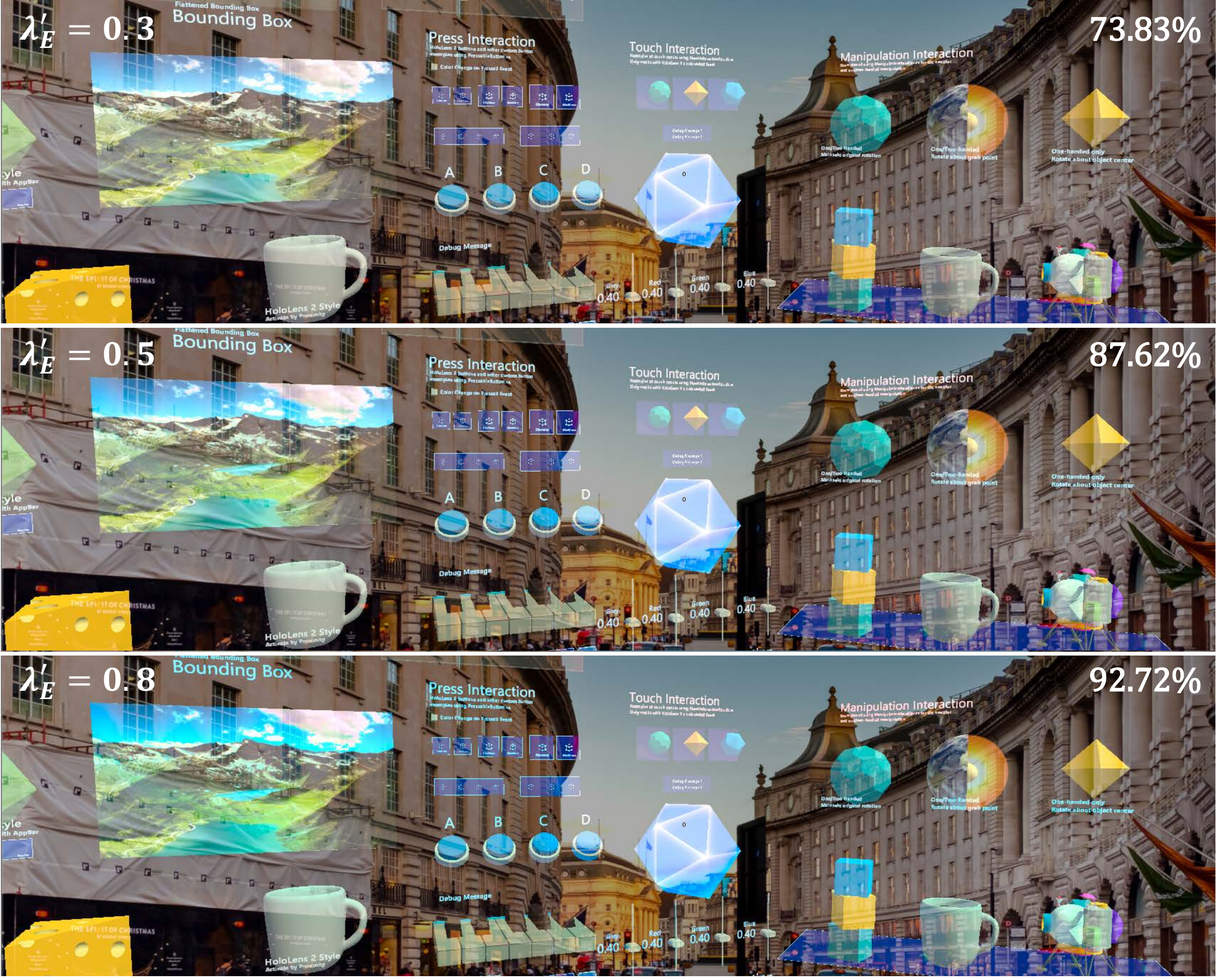}
  \caption{Results of different $\lambda^\prime_E$ for the same background in the simulated environment. Numbers indicate the percentage of enhanced pixels in all foreground pixels. One can see that the white text in front of yellow backgrounds looks bluish, whereas that in front of blue backgrounds looks yellowish. Please refer to the supplementary video for an example of different $\lambda^\prime_E$ captured by the HoloLens.}
  \label{fig:difference}
  \vspace{-0.6cm}
\end{figure}

We used different types of images as the background of the simulated environment to evaluate our method in various practical scenarios. Figure~\ref{fig:Results} shows some of the results. Virtual objects and dimmed background are blended directly \zyj{as~\eqref{eq:simplify}} instead of mixed by alpha blending to simulate the optical properties of OST-HMDs. Pixels that meet the following condition are marked as cyan in the last row of Figure~\ref{fig:Results}, meaning that these pixels have an increased color difference $\Delta E^\ast_{ab}$ from the corresponding background in perception: 
\begin{align}\label{eq:test_diff}
  \left(\Delta E^\ast_{ab}(l_{b},l_{opt})>\Delta E^\ast_{ab}(l_{b},l_{d})\right)\cap\left(\Delta E^\ast_{ab}(l_{opt},l_{d})\geq \lambda_{J\!N\!D}\right).
\end{align}
Here, $\lambda_{J\!N\!D}$ is about $2.3$ in the LAB color space \cite{Mahy1994}. Note that not all virtual content can gain perceptual color contrast enhancement. Given the constraints mentioned above, the color difference between certain optimal colors and their original displayed colors is less than JND. Moreover, some display colors are initially the complementary colors of the background colors and do not require further enhancement. To better demonstrate the effectiveness of our approach, we validated it with varying $\lambda^\prime_E$ with the same background, as shown in Figure~\ref{fig:difference}. Generally, larger color difference leads to an increase of the enhanced pixels in quantities and makes the displayed color more complementary to the background color. 

\begin{figure}
  \centering 
  \includegraphics[width=\columnwidth]{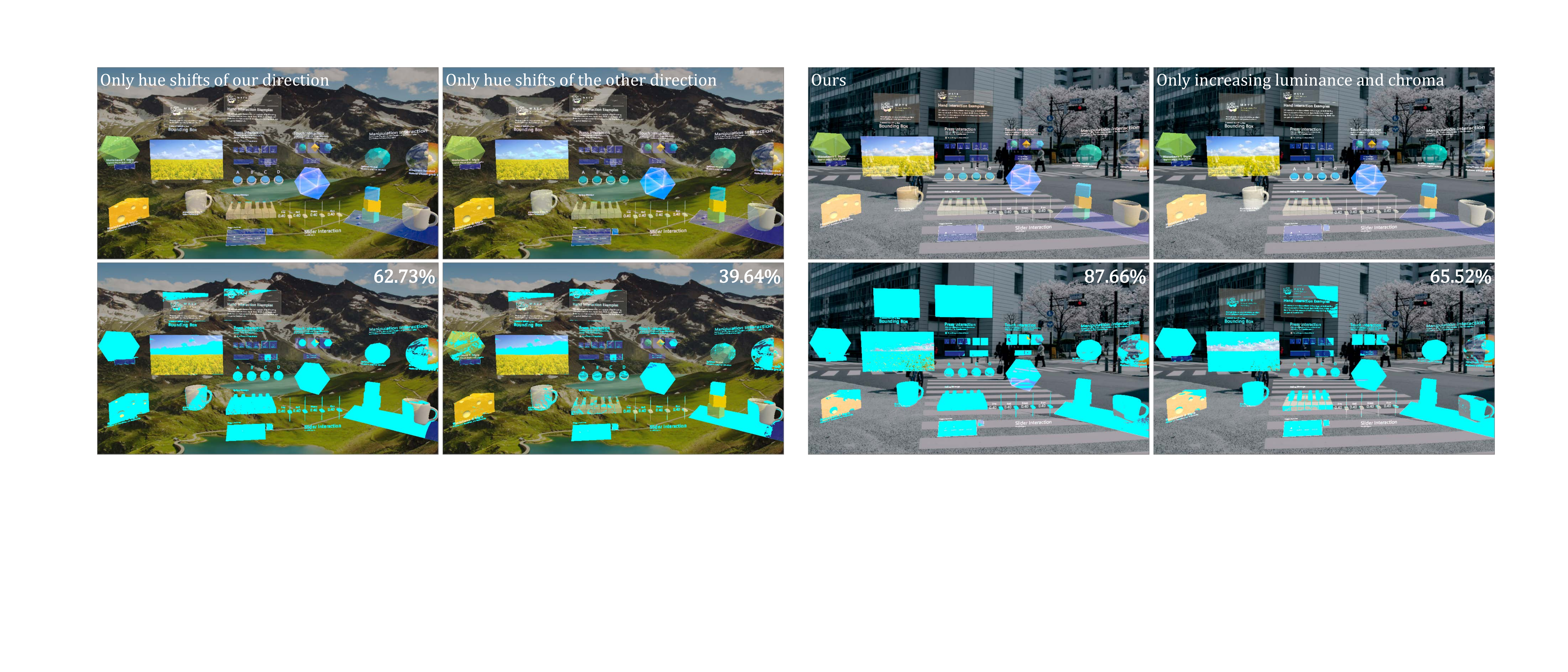}
  \caption{Results of different enhancements in the simulated environment, with the same change in color difference. \textbf{Left:} Our color contrast enhancement. \textbf{Right:} Enhancing color contrast by increasing luminance and chroma only. The numbers in figures indicate the corresponding percentage of the enhanced pixels.}
  \label{fig:HueChromaComparison}
\end{figure}

\begin{figure}
  \centering
  \subfloat[]{
  \includegraphics[width=0.475\columnwidth]{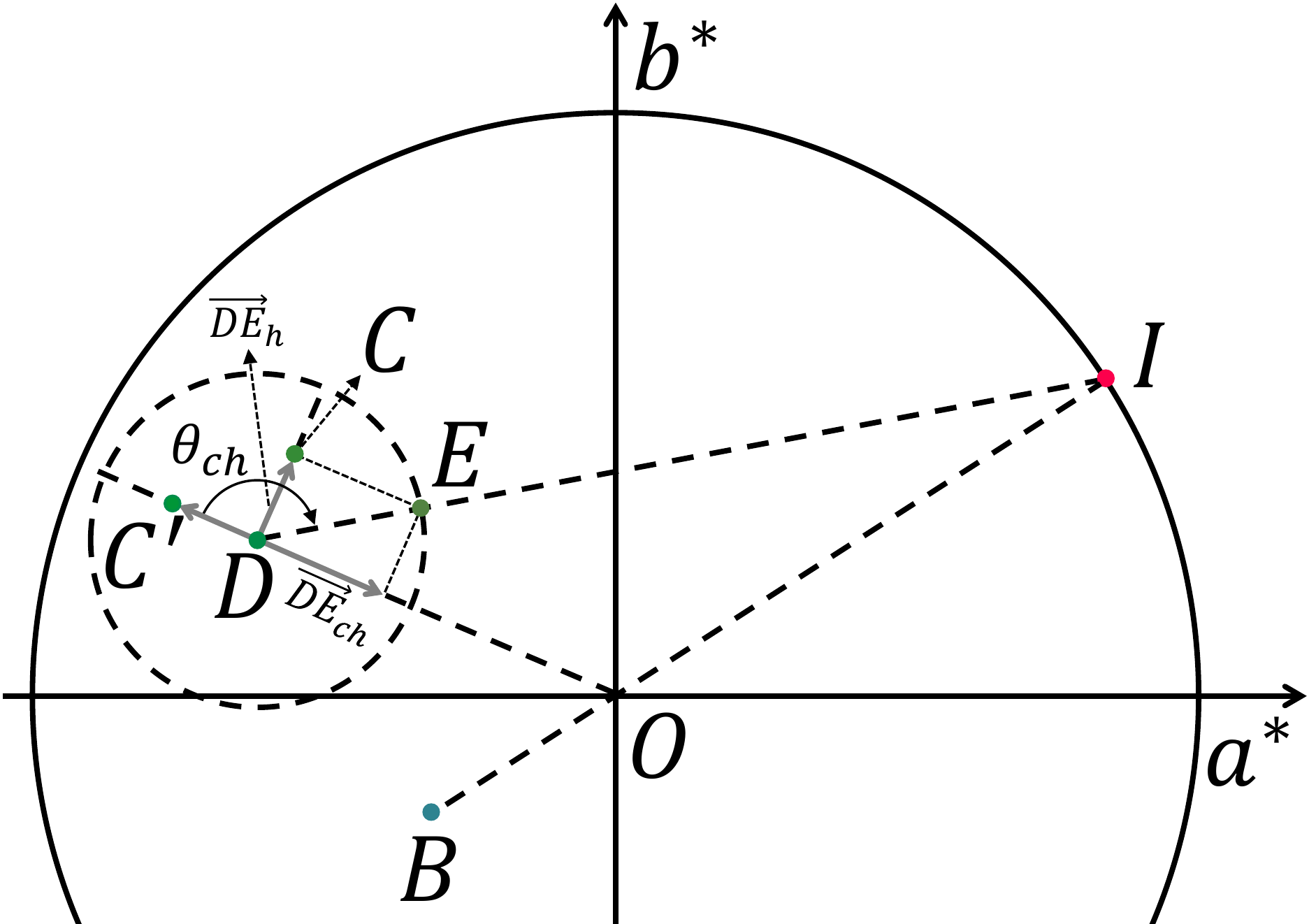}
  \label{fig:C_prime}
  }
  \subfloat[]{
  \includegraphics[width=0.475\columnwidth]{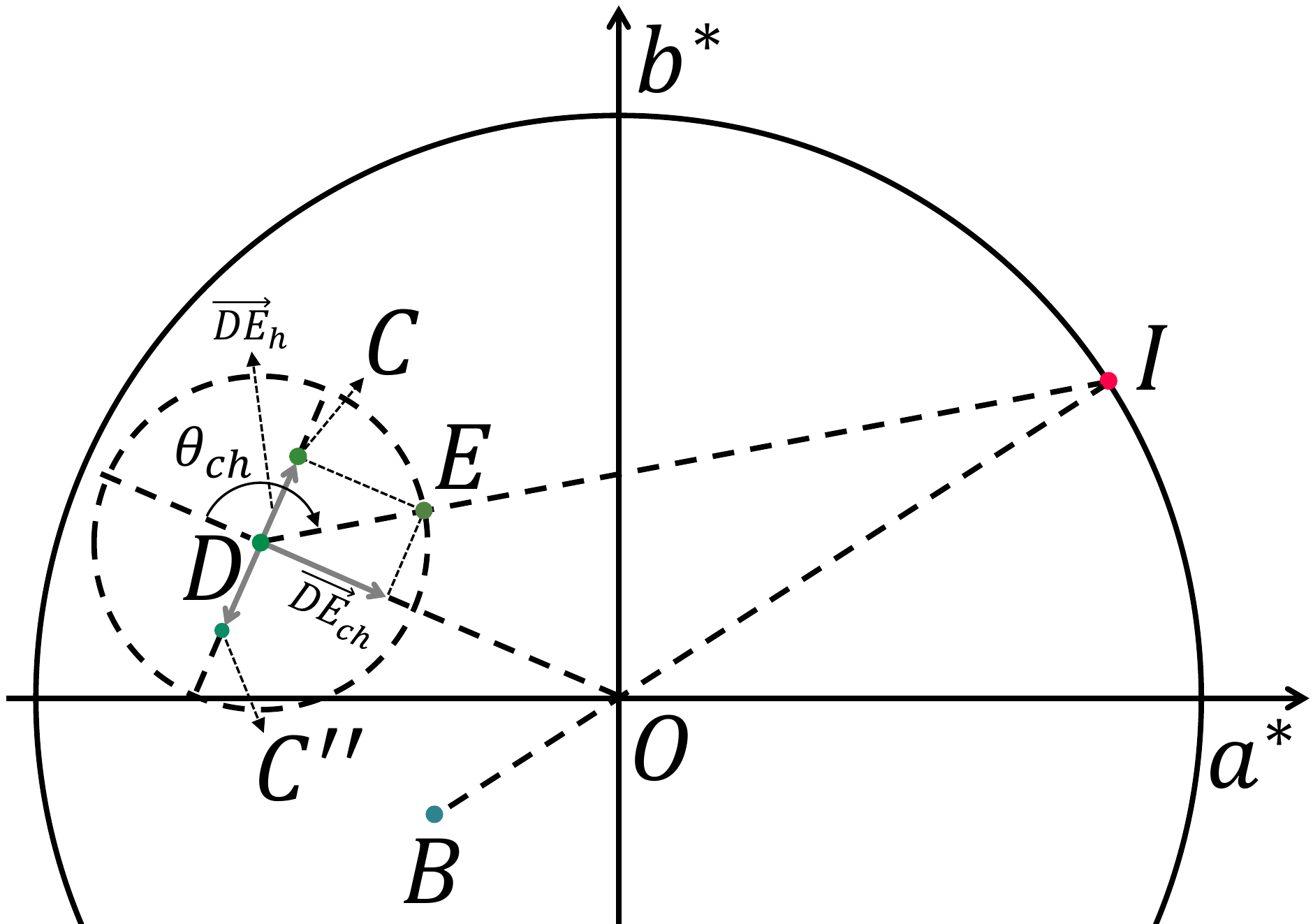}
  \label{fig:C_primeprime}
  }
  \caption{
    (\textbf{a}) A 2D illustration of increasing luminance and chroma only. $C^{\prime}$ is the coordinates of optimized color following this enhancement. The length of $\protect\overrightarrow{DC^{\prime}}$ is equal to $\protect\overrightarrow{DE}_{h}$, and its direction is along $\protect\overrightarrow{OD}$. 
    (\textbf{b}) A 2D illustration of enhancing color contrast in another direction of the hue shift. $C^{\prime \prime}$ is the coordinates of optimized color following this enhancement. The length of $\protect\overrightarrow{DC^{\prime \prime}}$ is equal to $\protect\overrightarrow{DE}_{h}$, while its direction is opposite to $\protect\overrightarrow{DE}_{h}$.
  }
  \vspace{-0.6cm}
\end{figure}

\zyj{To evaluate the enhancement contributed by hues, we compared results (Figure~\ref{fig:HueChromaComparison}) produced by our approach and just increasing luminance and chroma, with the same $\lambda^\prime_E$ of $0.4$. Figure~\ref{fig:C_prime} shows a two-dimensional illustration of this enhancement. The changes in color difference of each pixel between the two enhancements are identical. Pixels that have an increased perceptual color difference are marked as cyan like Figure~\ref{fig:Results}. In most cases, approach considering hues produces more enhancement than increasing luminance and chroma only. Some virtual objects like the coffee cup in Figure~\ref{fig:HueChromaComparison} are indeed more visible by applying the latter approach, but texture details like shadows are distorted.}

\begin{figure}
  \centering 
  \includegraphics[width=\columnwidth]{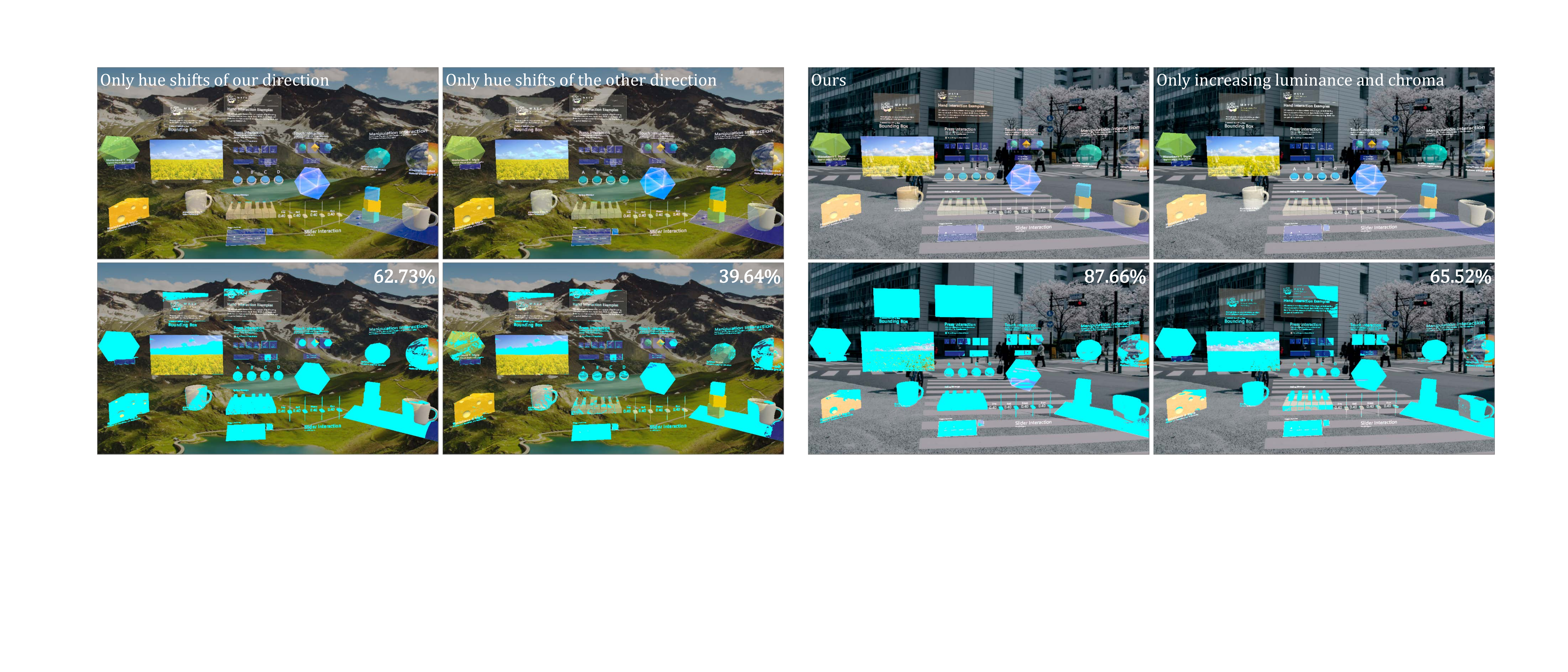}
  \caption{Results of different hue shifts in the simulated environment, with the same control conditions (i.e., iso-color-difference, iso-luminance, and iso-chroma). \textbf{Left:} Color contrast enhancement produced only by hue shifts of our method. \textbf{Right:} Another enhancement produced only by hue shifts but in a different direction.  Numbers in figures denote the corresponding percentage of the enhanced pixels in all foreground pixels.}
  \label{fig:otherHueDir}
  \vspace{-0.6cm}
\end{figure}

\zyj{The effects of color contrast enhancement produced by two directions of the hue shift were also evaluated. Figure~\ref{fig:otherHueDir} shows a comparison result, where $\lambda^\prime_E$ was both set to $0.4$. Figure~\ref{fig:C_primeprime} presents an illustration of the two directions. The first direction is based on our method, and the other direction has the same control conditions. The optimized colors of the two enhancements are the same as the original colors in luminance. The changes in chroma and color difference between the two enhancements are also identical. In fact, there is only one other direction ($\protect\overrightarrow{DC^{\prime \prime}}$) that is iso-color-difference, iso-luminance, and iso-chroma compared to $\protect\overrightarrow{DC}$. Pixels have an increased perceptual color difference are marked as cyan. Normally, A considerable degree of color contrast can be enhanced by hue shifts only. Moreover, the enhancement along the hue shift direction we proposed is better than that along the other hue shift direction under the same conditions.}

We also evaluated our algorithm on the HoloLens. Figure~\ref{fig:teaser} shows one of the experimental results, where the scaled color difference threshold $\lambda^\prime_E$ we used was $0.4$. In front of a yellow background, our method shifts the displayed color to the complementary direction (blue) of the background color, to enhance the color difference for better visual distinctions. For example, the sky and the ground of the landscape photograph look bluish, as well as the text in the scene. However, subject to the aforementioned constraints (Subsection \ref{ssc:constraints}), the chroma of yellow cheese and mantle does not decrease. Overall, our method is able to keep the consistency from the original displayed color.

\begin{figure}
  \centering 
  \includegraphics[width=\columnwidth]{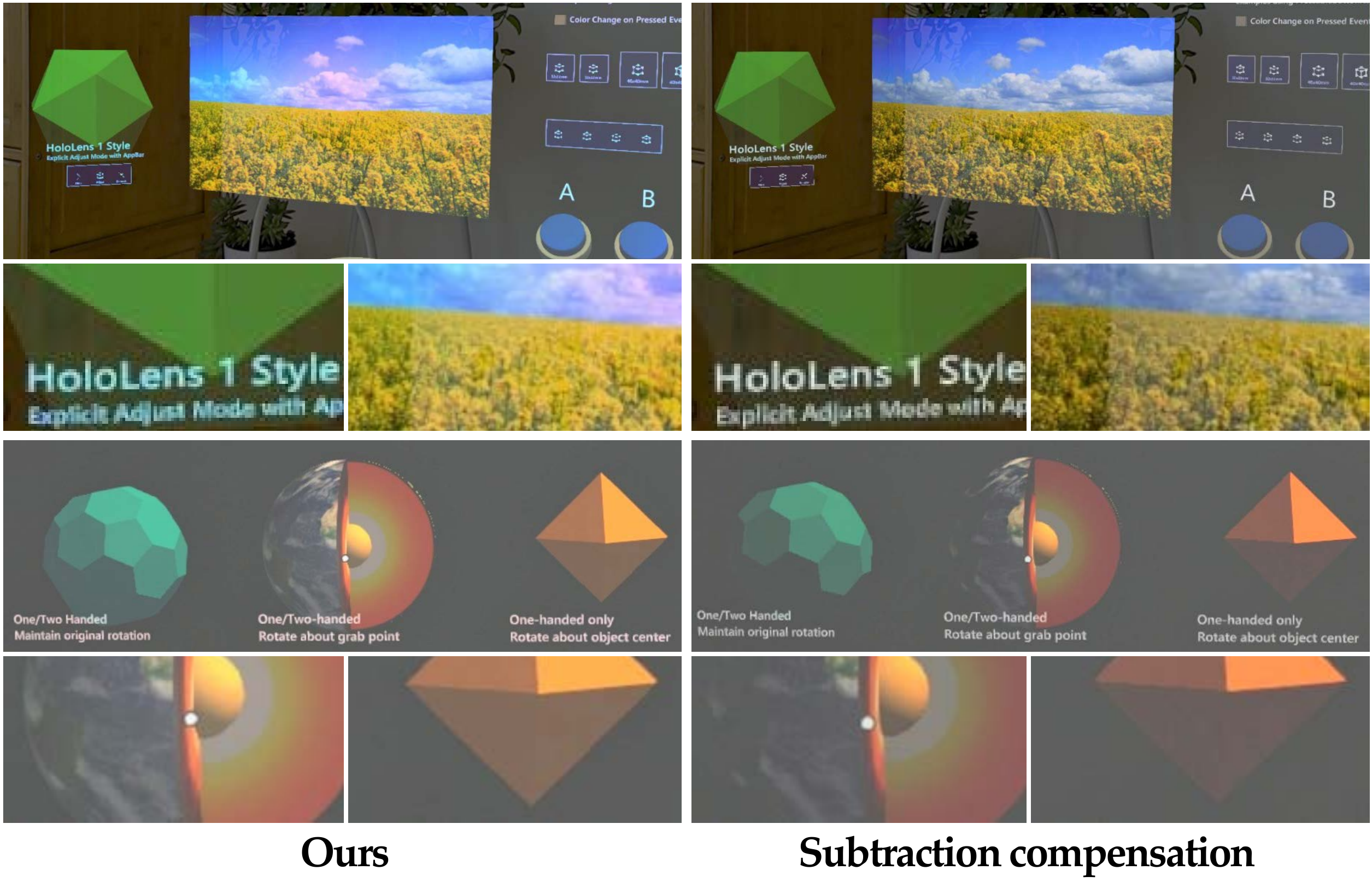}
  \caption{Rendering results of the two methods on the HoloLens, in front of different backgrounds. \textbf{Left}: Our color contrast enhancement ($\lambda^\prime_E = 0.4$). \textbf{Right:} Subtraction compensation \cite{Weiland2009} ($k_v=1.0$ and $k_b=0.4$, \zyj{where $k_v=1.0$ means the same display color in both methods, and $k_b=0.4$ denotes $40\%$ lens transparency, i.e., $60\%$ attenuation of background luminance).} When the background is light gray (such as a white wall), our method has limited optimization for rendered pixels. By contrast, the subtraction compensation causes degradation, making the virtual contents more transparent. Please refer to the supplementary video for another comparison result between the two methods.}
  \label{fig:algorithmComparison}
  \vspace{-0.2cm}
\end{figure}

\begin{figure}[t]
  \centering 
  \includegraphics[width=\columnwidth]{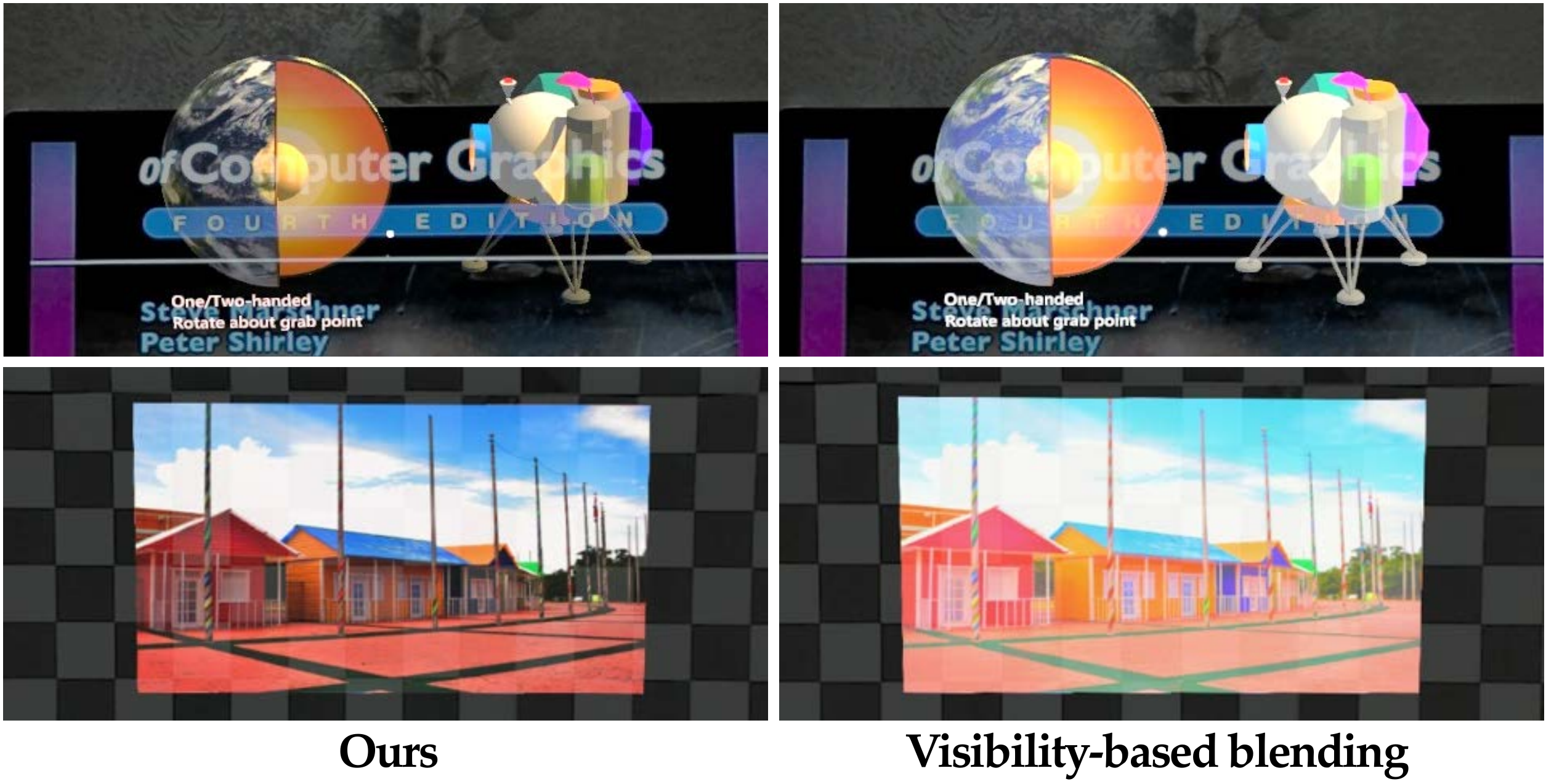}
  \caption{Rendering results of the two methods on the HoloLens, in front of different backgrounds. \textbf{Left}: Our color contrast enhancement ($\lambda^\prime_E = 0.4$). \textbf{Right:} Visibility-enhanced blending \cite{Fukiage2014} ($V_t = 1.5$ \zyj{, which is given by the authors in their paper}). When the background is an achromatic pattern, our method has little optimization for rendered pixels. By contrast, visibility-enhanced blending improves the distinctions between virtual contents and physical background but decreases contrast within surfaces of objects and leads to a possibly unintended change in color. Please refer to the supplementary video for another comparison result between the two methods.}
  \label{fig:algorithmComparison2}
  \vspace{-0.6cm}
\end{figure}

Although our goal is to improve the distinctions between virtual contents and surrounding backgrounds rather than correcting colors, we compared our color contrast enhancement with a typical compensation method \cite{Weiland2009}. Figure~\ref{fig:algorithmComparison} presents the experimental results. \zyj{Note that this result of our enhancement shows the maximum color contrast under a given $\lambda^\prime_E$. If one needs less contrast, but with more consistency with the original color, the threshold $\lambda^\prime_E$ is tweakable.} As noted previously, subtraction compensation reduces the luminance of rendered pixels, leading to low visual distinctions between virtual objects and the physical environment. Additionally, we compared our method with the visibility-based blending approach \cite{Fukiage2014}, as shown in Figure~\ref{fig:algorithmComparison2}. This blending method increases the brightness at the expense of contrast within surfaces, resulting in a washed-out visual effect of the virtual content.

\subsection{User Study I: Threshold Preference}
\label{ssc:user_study_1} 

\begin{figure}
  \centering 
  \includegraphics[width=\columnwidth]{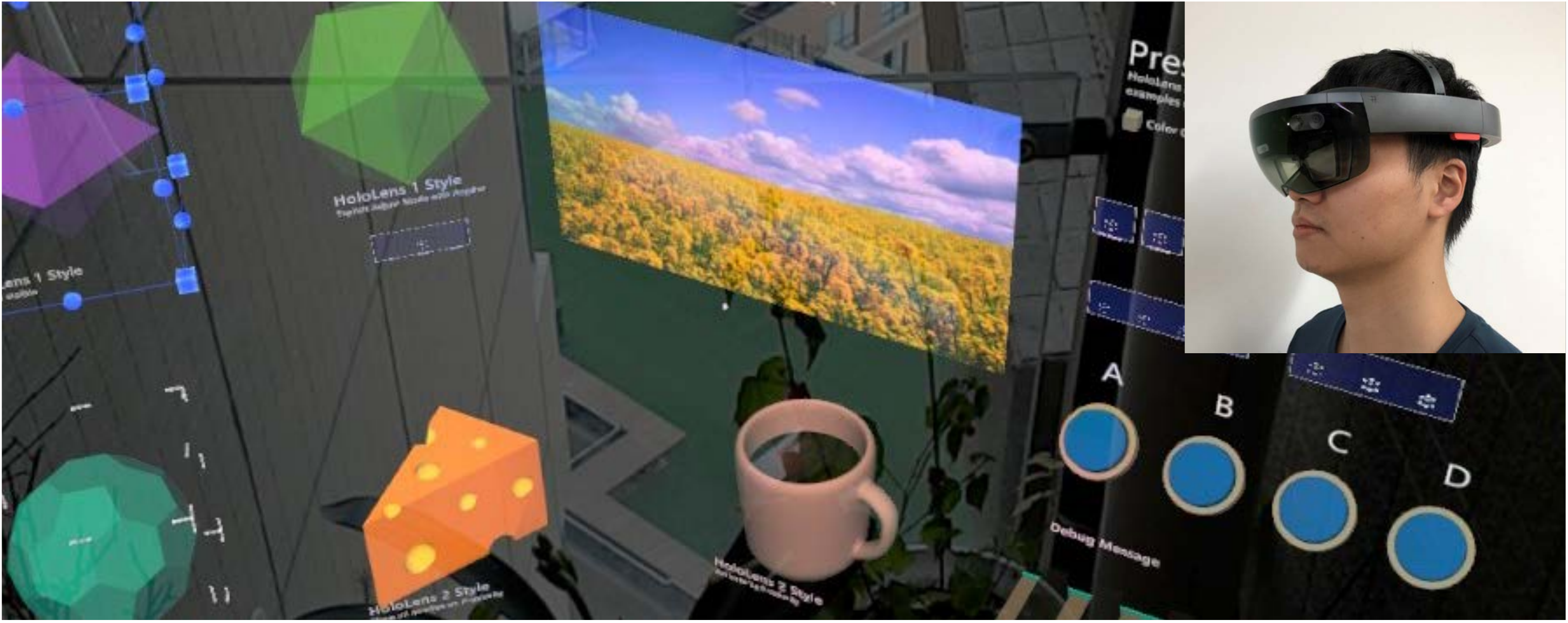}
  \caption{Scene perceived from the HoloLens by participants in a daily environment, with our color contrast enhancement enabled ($\lambda^\prime_E = 0.4$). One can see that the photograph in front of the green background looks reddish.}
  \label{fig:UserPerspective}
  \vspace{-0.6cm}
\end{figure}

We conducted three user studies to evaluate our color contrast enhanced rendering subjectively. We first performed a two-alternative forced choice (2AFC) subjective experiment, wherein the participants were asked to compare the results of color-contrast-enhanced rendering based on our approach and original rendering. We provided five levels of $\lambda^\prime_E$, ranging from $0.2$–$1.0$, with a step of $0.2$. During the experiment, participants wore the HoloLens and explored the modified \textit{Hand Interaction Examples} scene freely in various environments (see Figure~\ref{fig:UserPerspective} for an example). \zyj{In this scene, dozens of virtual objects were placed surrounding the user. Owing to the small FoV of the HoloLens, participants needed to rotate their head (change the camera viewport) to see different virtual objects with different backgrounds.} Then, participants performed full comparisons by freely toggling between the two results shown in the HoloLens, and then were asked two questions. First, which one of the results is more distinguishable from surrounding backgrounds. Second, which result looks more natural. \zyj{We asked each participant to look at three randomly picked objects, where each object is with one camera viewport.} Thus, each participant compared results in three random camera viewports under five color differences and gave a total of $15$ choices for each question.

A total of 15 participants, 12 males and 3 females, with a mean age of $23.6$ years (range $21$–$28$), took part in the experiments. All participants had normal or corrected-to-normal vision without any form of color blindness. Participants gave informed consent to take part in this study. \zyj{Before beginning the study, we required each participant to perform an eye-to-display calibration through a built-in calibration application of the HoloLens.} Finally, we received $45$ choices of each question under each color difference ($\lambda^\prime_E$). The results show that our method is more distinguishable from surrounding backgrounds in $142$ of $180$ comparisons when $\lambda^\prime_E \geq 0.4$ ($p < 0.01$, Binom. test). In addition, when  $\lambda^\prime_E \geq 0.6$, the original rendering images are statistically more natural ($35$ of $135$, $p < 0.01$, Binom. test).

\begin{figure}
  \centering 
  \includegraphics[width=\columnwidth]{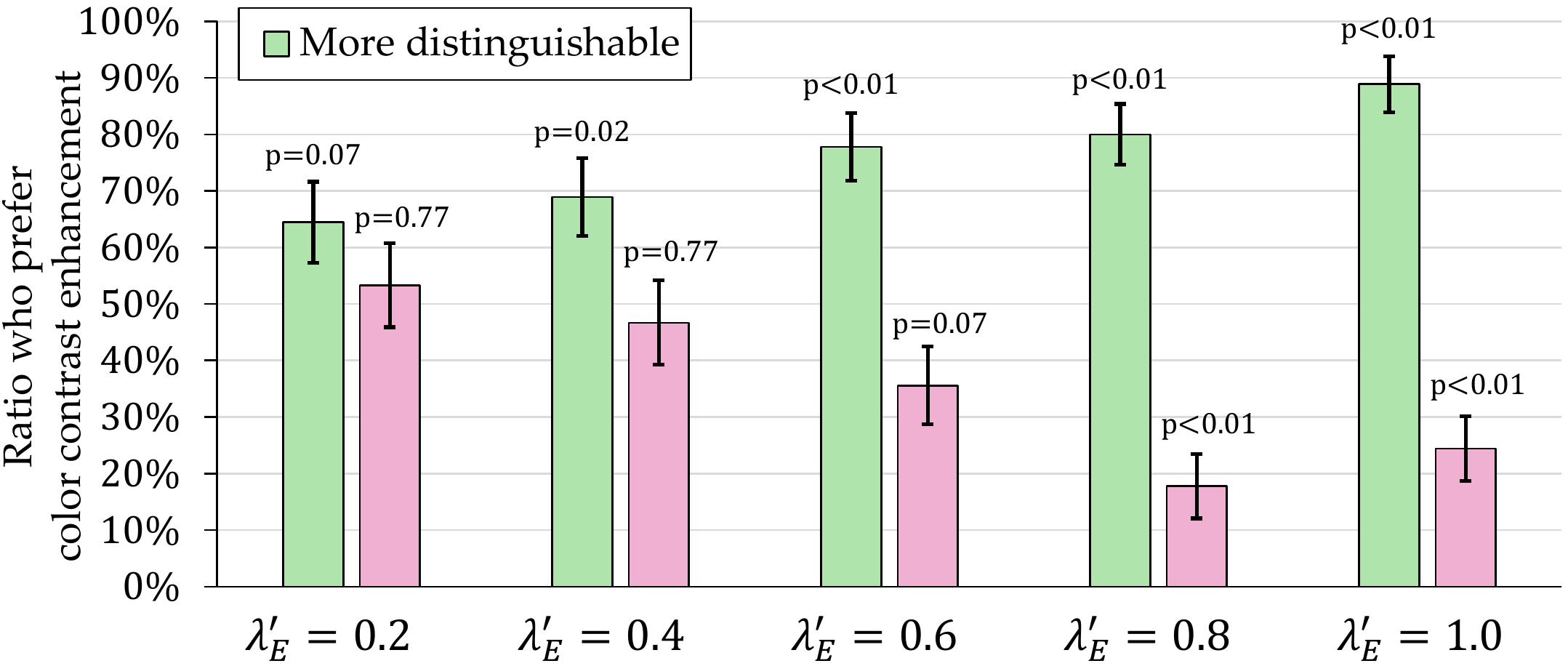}
  \caption{Results of our first subjective experiment that the participants compared our color-contrast-enhanced rendering with the original rendering. Participants' preferences are shown as percentages. $\lambda^\prime_E$ represents the scaled color difference threshold. p-value (Binom. test) is shown above the column, accordingly. The error bars represent standard error.}
  \label{fig:userstudyresults}
  \vspace{-0.4cm}
\end{figure}

Figure~\ref{fig:userstudyresults} shows the preferences of the participants. For all color differences tested on the HoloLens, our method is preferred more often than original rendering in distinction. On the other hand, the naturalness of color-contrast-enhanced images significantly decreases as $\lambda^\prime_E$ increases when $\lambda^\prime_E \geq 0.6$. \zyj{These results show that the value of $\lambda^\prime_E$ is positively correlated with the distinction but negatively correlated with the naturalness, which means that there is a trade-off between contrast and consistency.} However, when $\lambda^\prime_E = 0.4$, the optimized virtual content is more distinguishable from the background ($31$ of $45$, $p = 0.02$, Binom. test), whereas the difference of naturalness between the optimized and the original virtual content is not statistically significant ($21$ of $45$, $p = 0.77$, Binom. test). We believe that our approach has successfully found a trade-off between \zyj{contrast and consistency}.

\subsection{User Study II: Hue Evaluation}
\label{ssc:user_study_2}

\zyj{To subjectively evaluate the effect of hues, we performed another 2AFC experiment. Participants were asked to make two comparisons between results of 1) our color contrast enhancement and enhancing color contrast by increasing luminance and chroma only; 2) two different directions of hue shift (see Section~\ref{ssc:evaluation}). Participants wore the HoloLens and freely explored the same scene as User Study \uppercase\expandafter{\romannumeral1} in various environments. For each comparison, participants freely toggled between the two rendering results and then were asked one question that which of the results is more distinguishable from the surrounding background. Each participant compared the results of three random viewports and gave a choice for each viewport of each comparison. We fixed $\lambda^\prime_E$ to $0.4$.}

\zyj{Sixteen participants, including 4 females and 12 males with an average age of $24.6$ years old (range $23$–$28$), volunteered. All participants had normal or corrected-to-normal vision without any form of color blindness. Seven of them participated in the previous study. Participants gave informed consent to take part in this study. Before starting, we required each participant to perform the eye-to-display calibration.}

\begin{figure}
  \centering 
  \includegraphics[width=\columnwidth]{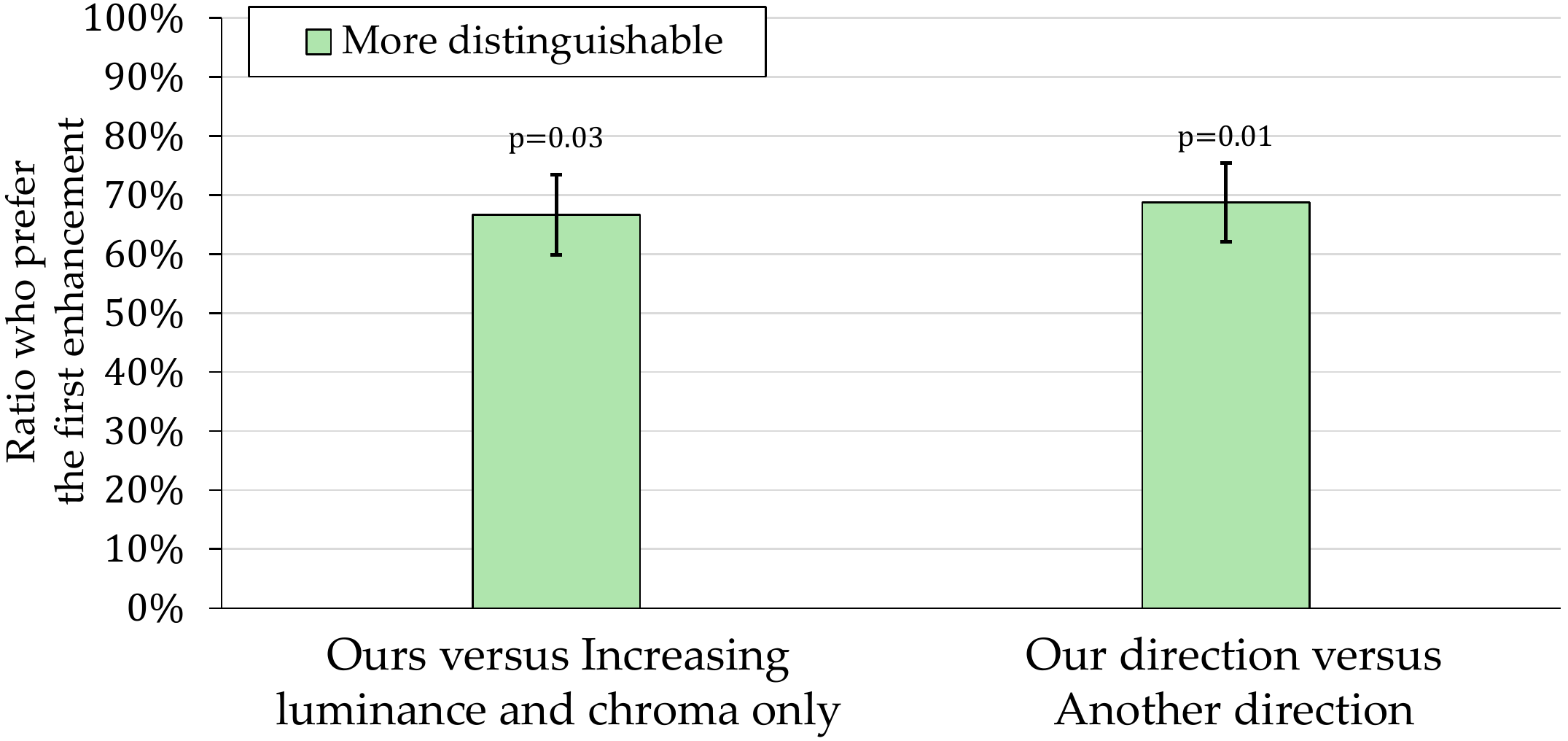}
  \caption{Results of our second subjective experiment that the participants compared the results of different enhancements. Participants' preferences are shown as percentages. p-value (Binom. test) is shown above the column, respectively. The error bars represent the standard error.}
  \label{fig:userstudyresults2}
  \vspace{-0.6cm}
\end{figure}

\zyj{For each comparison, a total of 48 choices was reported from participants. Figure~\ref{fig:userstudyresults2} shows the preference of the participants. For both comparisons, the first enhancement was preferred more often than the second one. Specifically, our color contrast enhancement is more distinguishable than increasing luminance and chroma only ($32$ of $48$, $p=0.03$, Binom. test). Also, the hue shift direction we proposed is more distinguishable than that along the other hue shift direction under the same conditions ($33$ of $48$, $p=0.01$, Binom. test). These results indicate that hue shifts in an appropriate direction can further improve the visual distinctions between rendered images and the surrounding environment.}

\subsection{User Study III: Method Comparison}
\label{ssc:user_study_3}

We conducted another 2AFC experiment in which the participants were asked to compare the results of our color contrast enhanced rendering with those of 1) subtraction compensation \cite{Weiland2009}; 2) visibility-enhanced blending \cite{Fukiage2014}. Similar to the previous user studies, participants wore the HoloLens and freely perceived virtual objects taken from the same scene, and then fully compared the rendered images of the two methods by freely toggling between them. For each related method, participants were asked three questions. First, which of the results is more distinguishable from surrounding backgrounds. Second, which result has higher contrast. Third, which result looks more natural. Each participant compared rendered images of three random viewports and gave six choices for each question. We fixed the $\lambda^\prime_E$ of our method to $0.4$.

A total of 12 participants consisting of 5 females and 7 males volunteered, ages $18$–$33$ (mean $22.6$). None of the participants exhibited signs of any form of color blindness. All participants had normal or corrected-to-normal vision. No participant took part in User Study \uppercase\expandafter{\romannumeral1} and \uppercase\expandafter{\romannumeral2}. Participants gave informed consent to participate in this study. \zyj{We required each participant to perform the eye-to-display calibration before starting the study. To further ensure the accuracy of the camera-to-display calibration, we also required the participants to be at least $100$ cm away from surrounding environments.}

\begin{figure}
  \centering 
  \includegraphics[width=\columnwidth]{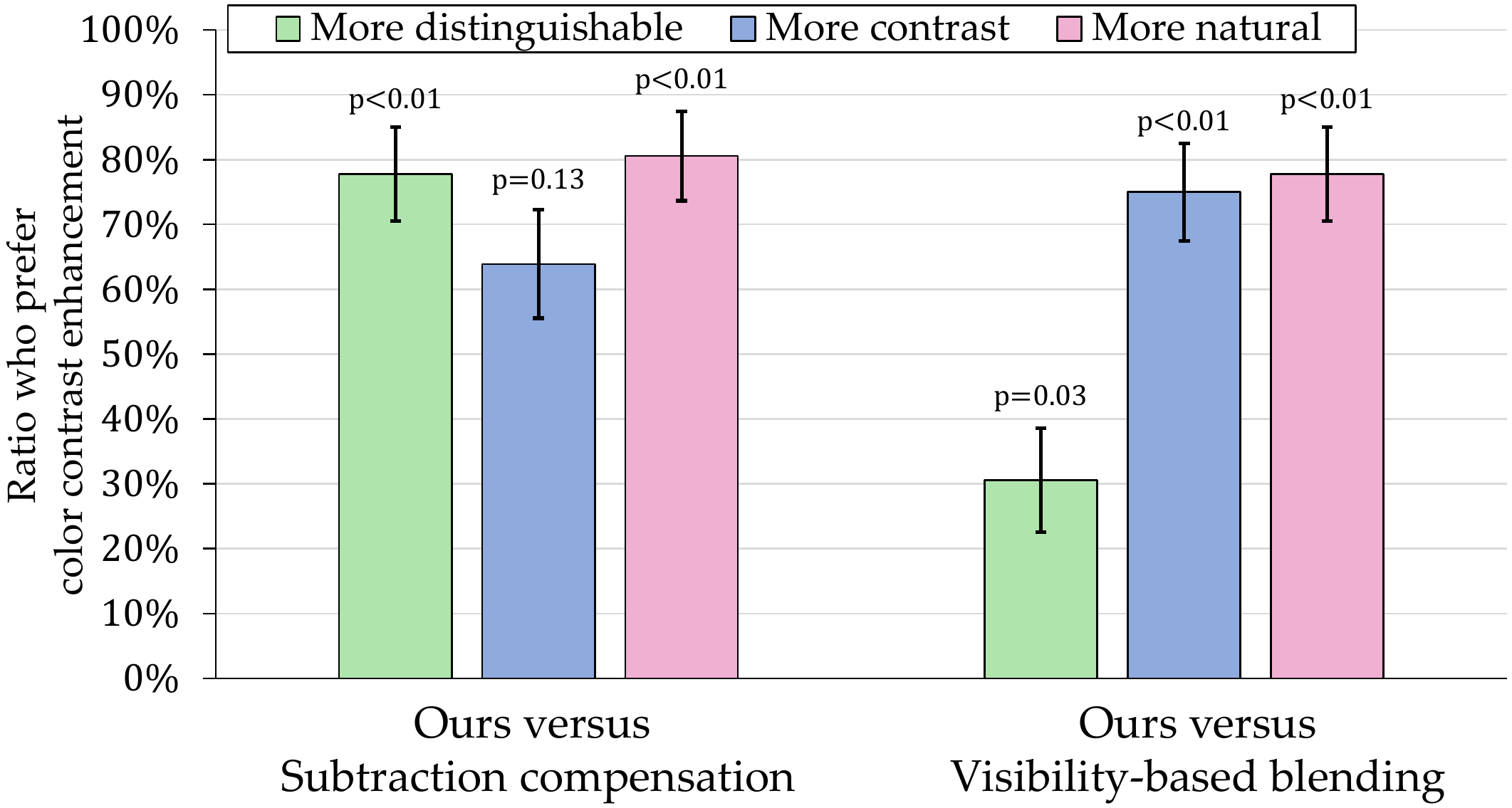}
  \caption{Results of our second subjective experiment that the participants compared our color-contrast-enhanced rendering with subtraction compensation \cite{Weiland2009} ($k_v=1$ and $k_b=0.4$) and visibility-enhanced blending \cite{Fukiage2014} ($V_t = 1.5$). Participants' preferences are shown as percentages. p-value (Binom. test) is shown above the column, respectively. The error bars represent standard error.}
  \label{fig:userstudyresults3}
  \vspace{-0.2cm}
\end{figure}

Finally, we collected 36 choices for every question of each compared method. Figure~\ref{fig:userstudyresults3} shows the preferences of the participants. For the first question, participants preferred our method more often than subtraction compensation ($28$ of $36$, $p<0.01$, Binom. test). However, our algorithm is less preferred than visibility-enhanced blending ($11$ of $36$, $p=0.03$, Binom. test). For the second question, our results are considered to have higher contrast compared with visibility-enhanced blending ($27$ of $36$, $p<0.01$, Binom. test). For the third question, the rendering images generated by our algorithm are statistically more natural than the other two methods ($57$ of $72$, $p<0.01$, Binom. test). These results mean that, in most cases, our approach is regarded as more distinguishable \zyj{and natural} than subtraction compensation, whereas the contrast \zyj{and naturalness} of rendered images are significantly higher than visibility-enhanced blending.

\zyj{We've compared with the subtraction compensation approach \cite{Weiland2009}. We note that more sophisticated methods based on subtraction compensation (e.g.,~\cite{Langlotz2016,Ryu2016}) also fail to keep the original visibility of virtual content on common OST-HMDs.}
\zyj{In addition, results show that participants regarded the visibility-enhanced blending \cite{Fukiage2014} as more distinguishable than our approach. This is because their method increases the luminance of virtual contents extensively. For applications that pay more attention to texture details, it may not be an optimal solution.}

\subsection{Performance}
\label{ssc:performance}

\begin{figure}
  \centering 
  \includegraphics[width=\columnwidth]{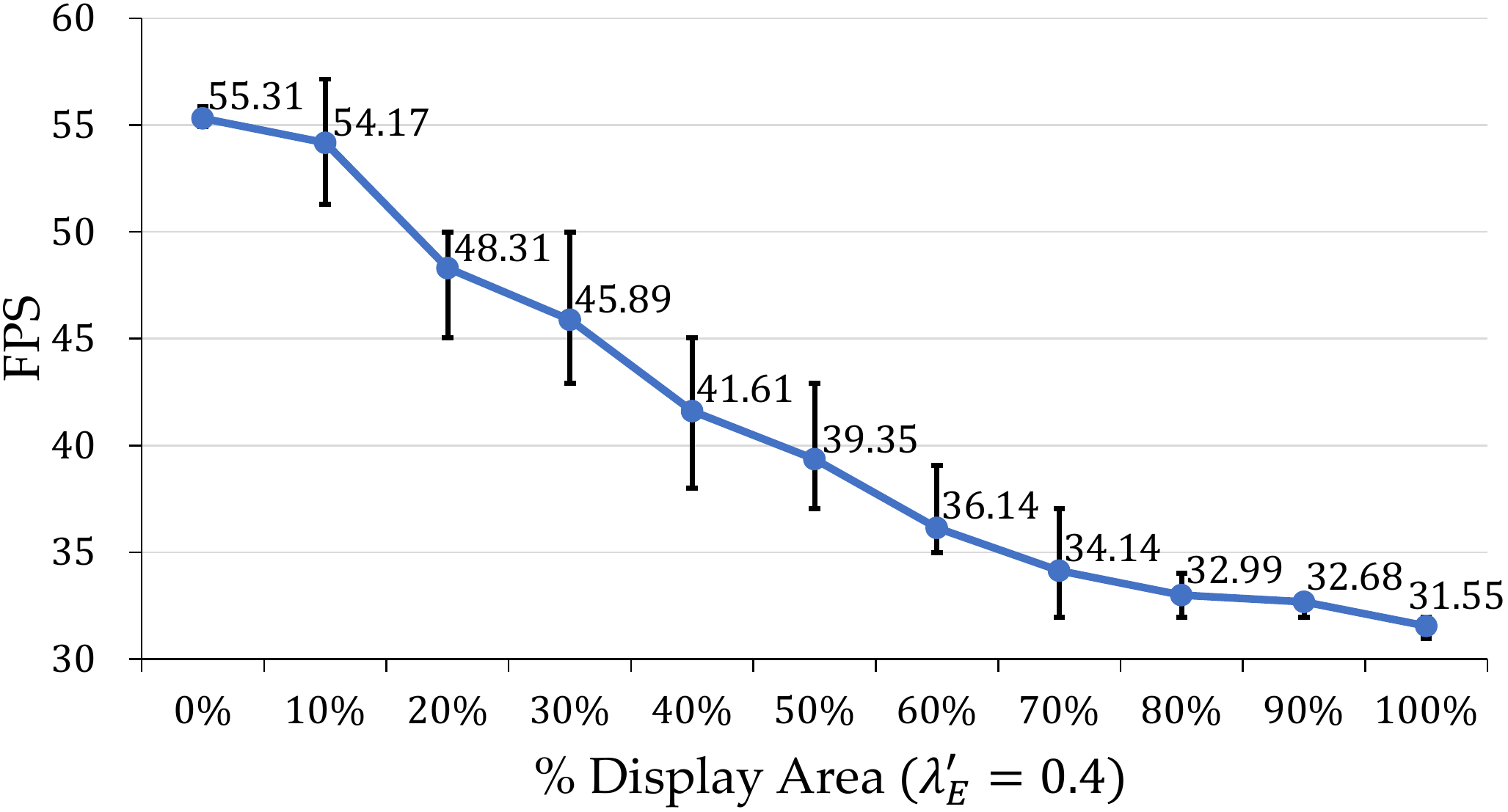}
  \caption{Performance of our color contrast enhancement as a full-screen post-processing effect on the HoloLens, measured in FPS. The error bars represent the maximum and minimum FPS.}
  \label{fig:performance}
  \vspace{-0.6cm}
\end{figure}

Our method does not rely on any precomputation. Therefore, it supports real-time rendering under various scenarios with dynamic camera viewports. Parameters such as color difference can be tuned in real-time. We measured the runtime performance on the HoloLens, which has a display area of $1268\times 720$ pixels for each eye. We rendered displayed contents at different viewports in the \textit{Hand Interaction Examples} scene, covering percentages of the display area that ranges from $0\%$–$100\%$, with a step of $10\%$. Different surrounding backgrounds are also used in our measurement. For each percentage, we sampled ten times and calculated the average FPS. Generally, reported FPS from the HoloLens varies from $31$–$55$, depending on the number of rendered pixels, as shown in Figure~\ref{fig:performance}.

\section{Limitation}
\label{sc:limitation}

Our approach, though not free from limitations, constitutes a promising avenue to future work. Although the experimental results indicate that our color contrast enhanced rendering for OST-HMDs works well in a proper threshold of color difference, finding an adaptive threshold corresponding to the current physical environment remains an unexplored valuable topic. 

There are several aspects in our implementation requiring elaboration. First, our method currently does not consider achromaticity. Although uncommon, it is reasonable to assume that the physical environment is completely achromatic (see bottom of Figure~\ref{fig:algorithmComparison} and \ref{fig:algorithmComparison2}). One possibility to incorporate this achromaticity situation into our optimization would be to increase the chroma of displayed colors. Second, the luminance of some optimal colors may be higher than the original displayed colors, resulting in a potential increase regarding power consumption of displays. In battery-powered OST-HMDs, battery life is one of the critical factors. Appropriate solutions that take the display brightness into account are worth exploring in future work.
\zyj{Note that for those applications weighting more about the geometry over color and texture, the scope of current work does not fit well. We consider this as an orthogonal problem and save it for future work.}

\section{Conclusion}
\label{sc:conclusion}

On the basis of a novel insight for color blending, we present an end-to-end, real-time color contrast enhancement algorithm for OST-HMDs that takes both chromaticity and luminance of displayed colors into account. Existing methods focus solely on changing the luminance of displays or environments to improve the distinction between virtual objects and real background, or compensating displayed colors to approximate the color originally intended. In this work, we further consider the impact of chromaticity to improve the color perception of rendered images from the perspective of \zyj{color contrast}. Specifically, we use the complementary color of the background as the search direction in the LAB color space to determine the optimal color of the display pixel under various constraints. 

To summarize, we strive to inspire future studies on color contrast enhanced rendering. Our current implementation achieves several key technical characteristics. First, it adaptively enhances the visibility of virtual objects. Second, it is real-time. Third, there is no need for any hardware change. Our algorithm is implemented on the GPU using pixel shaders, allowing users to tweak the parameters, such as the color difference, in real-time. We present results in simulated and real environments. In addition, both the objective evaluation and user study confirm that our approach can improve the perceptual contrast between displayed content and physical background in OST-HMDs, making the virtual objects more distinguishable from surrounding backgrounds \zyj{whereas achieving the maximum level of consistency with the original displayed color}.



\normalem
\bibliographystyle{IEEEtran}
\bibliography{template}

\begin{thebibliography}{10}
\providecommand{\url}[1]{#1}
\csname url@samestyle\endcsname
\providecommand{\newblock}{\relax}
\providecommand{\bibinfo}[2]{#2}
\providecommand{\BIBentrySTDinterwordspacing}{\spaceskip=0pt\relax}
\providecommand{\BIBentryALTinterwordstretchfactor}{4}
\providecommand{\BIBentryALTinterwordspacing}{\spaceskip=\fontdimen2\font plus
\BIBentryALTinterwordstretchfactor\fontdimen3\font minus
  \fontdimen4\font\relax}
\providecommand{\BIBforeignlanguage}[2]{{%
\expandafter\ifx\csname l@#1\endcsname\relax
\typeout{** WARNING: IEEEtran.bst: No hyphenation pattern has been}%
\typeout{** loaded for the language `#1'. Using the pattern for}%
\typeout{** the default language instead.}%
\else
\language=\csname l@#1\endcsname
\fi
#2}}
\providecommand{\BIBdecl}{\relax}
\BIBdecl

\bibitem{Cakmakci2004}
O.~Cakmakci, Y.~Ha, and J.~P. Rolland, ``A compact optical see-through
  head-worn display with occlusion support,'' in \emph{Proceedings of the 3rd
  IEEE/ACM International Symposium on Mixed and Augmented Reality}, ser. ISMAR
  '04.\hskip 1em plus 0.5em minus 0.4em\relax Washington, DC, USA: IEEE
  Computer Society, 2004, pp. 16--25.

\bibitem{Kiyokawa2000}
K.~{Kiyokawa}, Y.~{Kurata}, and H.~{Ohno}, ``An optical see-through display for
  mutual occlusion of real and virtual environments,'' in \emph{Proceedings
  IEEE and ACM International Symposium on Augmented Reality (ISAR)}, Oct 2000,
  pp. 60--67.

\bibitem{Gao2012}
C.~{Gao}, Y.~{Lin}, and H.~{Hua}, ``Occlusion capable optical see-through
  head-mounted display using freeform optics,'' in \emph{2012 IEEE
  International Symposium on Mixed and Augmented Reality (ISMAR)}, Nov 2012,
  pp. 281--282.

\bibitem{Wilson2017}
A.~Wilson and H.~Hua, ``Design and prototype of an augmented reality display
  with per-pixel mutual occlusion capability,'' \emph{Opt. Express}, vol.~25,
  no.~24, pp. 30\,539--30\,549, 2017.

\bibitem{Weiland2009}
C.~Weiland, A.-K. Braun, and W.~Heiden, ``Colorimetric and photometric
  compensation for optical see-through displays,'' in \emph{Universal Access in
  Human-Computer Interaction. Intelligent and Ubiquitous Interaction
  Environments}, C.~Stephanidis, Ed.\hskip 1em plus 0.5em minus 0.4em\relax
  Berlin, Heidelberg: Springer Berlin Heidelberg, 2009, pp. 603--612.

\bibitem{Hincapie-Ramos2015}
J.~D. Hincapi{\'{e}}-Ramos, L.~Ivanchuk, S.~K. Sridharan, and P.~P. Irani,
  ``{S}mart{C}olor: {R}eal-time color and contrast correction for optical
  see-through head-mounted displays,'' \emph{IEEE Transactions on Visualization
  and Computer Graphics}, vol.~21, no.~12, pp. 1336--1348, 2015.

\bibitem{Langlotz2016}
T.~Langlotz, M.~Cook, and H.~Regenbrecht, ``{Real-time radiometric compensation
  for optical see-through head-mounted displays},'' \emph{IEEE Transactions on
  Visualization and Computer Graphics}, vol.~22, no.~11, pp. 2385--2394, 2016.

\bibitem{Kim2017}
J.~{Kim}, J.~{Ryu}, S.~{Ryu}, K.~{Lee}, and J.~{Kim}, ``{[POSTER] O}ptimizing
  background subtraction for {OST-HMD},'' in \emph{2017 IEEE International
  Symposium on Mixed and Augmented Reality (ISMAR-Adjunct)}, Oct 2017, pp.
  95--96.

\bibitem{Fukiage2014}
T.~{Fukiage}, T.~{Oishi}, and K.~{Ikeuchi}, ``Visibility-based blending for
  real-time applications,'' in \emph{2014 IEEE International Symposium on Mixed
  and Augmented Reality (ISMAR)}, Sep 2014, pp. 63--72.

\bibitem{wolfe2006sensation}
J.~M. Wolfe, K.~R. Kluender, D.~M. Levi, L.~M. Bartoshuk, R.~S. Herz, R.~L.
  Klatzky, S.~J. Lederman, and D.~M. Merfeld, \emph{{Sensation and perception,
  4th ed.}}\hskip 1em plus 0.5em minus 0.4em\relax Wolfe, Jeremy M.:
  jwolfe@partners.org: Sinauer Associates, 2015.

\bibitem{Gruber2010}
L.~{Gruber}, D.~{Kalkofen}, and D.~{Schmalstieg}, ``Color harmonization for
  augmented reality,'' in \emph{2010 IEEE International Symposium on Mixed and
  Augmented Reality}, Oct 2010, pp. 227--228.

\bibitem{Itoh2015}
Y.~Itoh and G.~Klinker, ``Vision enhancement: Defocus correction via optical
  see-through head-mounted displays,'' in \emph{Proceedings of the 6th
  Augmented Human International Conference}, ser. AH '15.\hskip 1em plus 0.5em
  minus 0.4em\relax New York, NY, USA: ACM, 2015, pp. 1--8.

\bibitem{Oshima2016}
K.~{Oshima}, K.~R. {Moser}, D.~C. {Rompapas}, J.~E. {Swan}, S.~{Ikeda},
  G.~{Yamamoto}, T.~{Taketomi}, C.~{Sandor}, and H.~{Kato}, ``{S}harp{V}iew:
  {I}mproved clarity of defocussed content on optical see-through head-mounted
  displays,'' in \emph{2016 IEEE Virtual Reality (VR)}, Mar 2016, pp. 253--254.

\bibitem{Menk2013}
C.~Menk and R.~Koch, ``{Truthful color reproduction in spatial augmented
  reality applications},'' \emph{IEEE Transactions on Visualization and
  Computer Graphics}, vol.~19, no.~2, pp. 236--248, 2013.

\bibitem{Itoh2015a}
Y.~Itoh, M.~Dzitsiuk, T.~Amano, and G.~Klinker, ``Semi-parametric color
  reproduction method for optical see-through head-mounted displays,''
  \emph{IEEE Transactions on Visualization and Computer Graphics}, vol.~21,
  no.~11, pp. 1269--1278, 2015.

\bibitem{Oskam2012}
T.~{Oskam}, A.~{Hornung}, R.~W. {Sumner}, and M.~{Gross}, ``Fast and stable
  color balancing for images and augmented reality,'' in \emph{2012 Second
  International Conference on 3D Imaging, Modeling, Processing, Visualization
  Transmission}, Oct 2012, pp. 49--56.

\bibitem{Wetzstein2010}
G.~Wetzstein, W.~Heidrich, and D.~Luebke, ``Optical image processing using
  light modulation displays,'' \emph{Computer Graphics Forum}, vol.~29, no.~6,
  pp. 1934--1944, 2010.

\bibitem{Mori2018}
S.~Mori, S.~Ikeda, A.~Plopski, and C.~Sandor, ``{B}right{V}iew: {I}ncreasing
  perceived brightness of optical see-through head-mounted displays through
  unnoticeable incident light reduction,'' \emph{25th IEEE Conference on
  Virtual Reality and 3D User Interfaces, VR 2018 - Proceedings}, pp. 251--258,
  2018.

\bibitem{Itoh2019}
Y.~{Itoh}, T.~{Langlotz}, D.~{Iwai}, K.~{Kiyokawa}, and T.~{Amano}, ``Light
  attenuation display: Subtractive see-through near-eye display via spatial
  color filtering,'' \emph{IEEE Transactions on Visualization and Computer
  Graphics}, vol.~25, no.~5, pp. 1951--1960, 2019.

\bibitem{Lee2018}
K.-H. Lee and J.-O. Kim, ``{Visibility enhancement via optimal two-piece gamma
  tone mapping for optical see-through displays under ambient light},''
  \emph{Optical Engineering}, vol.~57, no.~12, pp. 1 -- 13, 2018.

\bibitem{Rathinavel2019}
K.~Rathinavel, G.~Wetzstein, and H.~Fuchs, ``Varifocal occlusion-capable
  optical see-through augmented reality display based on focus-tunable
  optics,'' \emph{IEEE Transactions on Visualization and Computer Graphics},
  vol.~25, no.~11, pp. 3125--3134, 2019.

\bibitem{Maimone2013}
A.~{Maimone} and H.~{Fuchs}, ``Computational augmented reality eyeglasses,'' in
  \emph{2013 IEEE International Symposium on Mixed and Augmented Reality
  (ISMAR)}, Oct 2013, pp. 29--38.

\bibitem{Smithwick2014}
Q.~Y.~J. Smithwick, D.~Reetz, and L.~Smoot, ``{LCD masks for spatial augmented
  reality},'' in \emph{Stereoscopic Displays and Applications XXV}, A.~J.
  Woods, N.~S. Holliman, and G.~E. Favalora, Eds., vol. 9011, no. March 2014,
  Mar 2014, p. 90110O.

\bibitem{Rhodes2019}
T.~Rhodes, G.~Miller, Q.~Sun, D.~Ito, and L.-Y. Wei, ``A transparent display
  with per-pixel color and opacity control,'' in \emph{ACM SIGGRAPH 2019
  Emerging Technologies}, ser. SIGGRAPH '19.\hskip 1em plus 0.5em minus
  0.4em\relax New York, NY, USA: ACM, 2019, pp. 5:1--5:2.

\bibitem{Ryu2016}
J.~{Ryu}, J.~{Kim}, K.~{Lee}, and J.~{Kim}, ``Colorimetric background
  estimation for color blending reduction of {OST-HMD},'' in \emph{2016
  Asia-Pacific Signal and Information Processing Association Annual Summit and
  Conference (APSIPA)}, Dec 2016, pp. 1--4.

\bibitem{Brown1997}
R.~O. Brown and D.~I. MacLeod, ``{Color appearance depends on the variance of
  surround colors},'' \emph{Current Biology}, vol.~7, no.~11, pp. 844--849, Nov
  1997.

\bibitem{Webster2002}
M.~A. Webster, G.~Malkoc, A.~C. Bilson, and S.~M. Webster, ``{Color contrast
  and contextual influences on color appearance},'' \emph{Journal of Vision},
  vol.~2, no.~6, pp. 7--7, 11 2002.

\bibitem{Ekroll2013}
V.~Ekroll and F.~Faul, ``Transparency perception: the key to understanding
  simultaneous color contrast,'' \emph{J. Opt. Soc. Am. A}, vol.~30, no.~3, pp.
  342--352, Mar 2013.

\bibitem{Klauke2015}
S.~Klauke and T.~Wachtler, ``{``Tilt''} in color space: {H}ue changes induced
  by chromatic surrounds,'' \emph{Journal of Vision}, vol.~15, no.~13, pp.
  17--17, 09 2015.

\bibitem{Jameson1959}
D.~Jameson and L.~M. Hurvich, ``Perceived color and its dependence on focal,
  surrounding, and preceding stimulus variables,'' \emph{J. Opt. Soc. Am.},
  vol.~49, no.~9, pp. 890--898, Sep 1959.

\bibitem{Krauskopf1986}
J.~Krauskopf, Q.~Zaidi, and M.~B. Mandlert, ``Mechanisms of simultaneous color
  induction,'' \emph{J. Opt. Soc. Am. A}, vol.~3, no.~10, pp. 1752--1757, Oct
  1986.

\bibitem{Jameson1961}
D.~Jameson and L.~M. Hurvich, ``Opponent chromatic induction: Experimental
  evaluation and theoretical account,'' \emph{J. Opt. Soc. Am.}, vol.~51,
  no.~1, pp. 46--53, Jan 1961.

\bibitem{Ekroll2012}
V.~Ekroll and F.~Faul, ``New laws of simultaneous contrast?'' \emph{Seeing and
  Perceiving}, vol.~25, no.~2, pp. 107 -- 141, 01 Jan. 2012.

\bibitem{Ratnasingam2017}
S.~Ratnasingam and B.~L. Anderson, ``{What predicts the strength of
  simultaneous color contrast?}'' \emph{Journal of Vision}, vol.~17, no.~2, pp.
  13--13, 02 2017.

\bibitem{Gabbard2010}
J.~L. {Gabbard}, J.~E. {Swan}, J.~{Zedlitz}, and W.~W. {Winchester}, ``More
  than meets the eye: {A}n engineering study to empirically examine the
  blending of real and virtual color spaces,'' in \emph{2010 IEEE Virtual
  Reality Conference (VR)}, Mar 2010, pp. 79--86.

\bibitem{Campbell1968}
F.~W. Campbell and J.~G. Robson, ``Application of fourier analysis to the
  visibility of gratings,'' \emph{The Journal of Physiology}, vol. 197, no.~3,
  pp. 551--566, Aug 1968.

\bibitem{Anderson1991}
S.~J. Anderson, K.~T. Mullen, and R.~F. Hess, ``Human peripheral spatial
  resolution for achromatic and chromatic stimuli: limits imposed by optical
  and retinal factors.'' \emph{The Journal of Physiology}, vol. 442, no.~1, pp.
  47--64, 1991.

\bibitem{Kress2017}
B.~C. Kress and W.~J. Cummings, ``11-1: {I}nvited {P}aper: {T}owards the
  ultimate mixed reality experience: {H}olo{L}ens display architecture
  choices,'' \emph{SID Symposium Digest of Technical Papers}, vol.~48, no.~1,
  pp. 127--131, 2017.

\bibitem{Mahy1994}
M.~Mahy, L.~Van~Eycken, and A.~Oosterlinck, ``Evaluation of uniform color
  spaces developed after the adoption of {CIELAB} and {CIELUV},'' \emph{Color
  Research \& Application}, vol.~19, no.~2, pp. 105--121, 1994.

\end{thebibliography}

\end{document}